\input lanlmac
\overfullrule=0pt
\parindent 25pt
\tolerance=10000
\input epsf

\newcount\figno
\figno=0
\def\fig#1#2#3{
\par\begingroup\parindent=0pt\leftskip=1cm\rightskip=1cm\parindent=0pt
\baselineskip=11pt
\global\advance\figno by 1
\midinsert
\epsfxsize=#3
\centerline{\epsfbox{#2}}
\vskip 12pt
{\bf Fig.\ \the\figno: } #1\par
\endinsert\endgroup\par
}
\def\figlabel#1{\xdef#1{\the\figno}}
\def\encadremath#1{\vbox{\hrule\hbox{\vrule\kern8pt\vbox{\kern8pt
\hbox{$\displaystyle #1$}\kern8pt}
\kern8pt\vrule}\hrule}}

\font\cmss=cmss10 \font\cmsss=cmss10 at 7pt

\def\Z{\relax\ifmmode\mathchoice
{\hbox{\cmss Z\kern-.4em Z}}{\hbox{\cmss Z\kern-.4em Z}}
{\lower.9pt\hbox{\cmsss Z\kern-.4em Z}}
{\lower1.2pt\hbox{\cmsss Z\kern-.4em Z}}\else{\cmss Z\kern-.4em
Z}\fi}

\font\cmss=cmss10
\font\cmsss=cmss10 at 7pt

\def\ep{{\epsilon}}

\def\a{{\alpha}}

\def\frac#1#2{{#1\over #2}}

\def\s{\sqrt}

\def\p{\partial}

\def\Re{{\rm Re}}

\def\al{\alpha'}
\def\de{\partial}

\def\lr{\leftrightarrow}
\def\f {\frac}
\def\ti{\tilde}

\def\ddd{\cdot\cdot\cdot}

\def\la{\langle}
\def\lb{\rangle}
\def\ep{\epsilon}

\def\ov{\overline}

\def\vp{\varphi}

\lref\MV{
J.~McGreevy and H.~Verlinde,
``Strings from tachyons: The c = 1 matrix reloaded,''
[arXiv:hep-th/0304224].}

\lref\KMS{
I.~R.~Klebanov, J.~Maldacena and N.~Seiberg,
``D-brane decay in two-dimensional string theory,''
[arXiv:hep-th/0305159].}

\lref\MTV{
J.~McGreevy, J.~Teschner and H.~Verlinde,
``Classical and quantum D-branes in 2D string theory,''
[arXiv:hep-th/0305194].
}

\lref\MoPlRa{
G.~W.~Moore, M.~R.~Plesser and S.~Ramgoolam,
``Exact S matrix for 2-D string theory,''
Nucl.\ Phys.\ B {\bf 377}, 143 (1992)
[arXiv:hep-th/9111035].
}

\lref\TT{T.~Takayanagi and N.~Toumbas,
``A matrix model dual of type 0B string theory in two dimensions,''
JHEP {\bf 0307}, 064 (2003)
[arXiv:hep-th/0307083].
}

\lref\DVV{R.~Dijkgraaf, H.~Verlinde and E.~Verlinde,
``String propagation in a black hole geometry,''
Nucl.\ Phys.\ B {\bf 371}, 269 (1992).
}

\lref\Kl{K.~Demeterfi, I.~R.~Klebanov and J.~P.~Rodrigues,
``The Exact S matrix of the deformed c = 1 matrix model,''
Phys.\ Rev.\ Lett.\  {\bf 71}, 3409 (1993)
[arXiv:hep-th/9308036].
}

\lref\KKK{V.~Kazakov, I.~K.~Kostov and D.~Kutasov,
``A matrix model for the two-dimensional black hole,''
Nucl.\ Phys.\ B {\bf 622}, 141 (2002)
[arXiv:hep-th/0101011].
}

\lref\KlR{I.~R.~Klebanov,
``String theory in two-dimensions,''
[arXiv:hep-th/9108019].}

\lref\six{M.~R.~Douglas, I.~R.~Klebanov, D.~Kutasov, J.~Maldacena,
E.~Martinec and N.~Seiberg,
``A new hat for the c = 1 matrix model,''
[arXiv:hep-th/0307195].
}

\lref\five{O.~DeWolfe, R.~Roiban, M.~Spradlin, A.~Volovich and J.~Walcher,
``On the S-matrix of type 0 string theory,''
[arXiv:hep-th/0309148].
}

\lref\FuHo{
T.~Fukuda and K.~Hosomichi,
``Super Liouville theory with boundary,''
Nucl.\ Phys.\ B {\bf 635}, 215 (2002)
[arXiv:hep-th/0202032].
}

\lref\BGV{N.~Berkovits, S.~Gukov and B.~C.~Vallilo,
``Superstrings in 2D backgrounds with R-R flux and new extremal
black  holes,''
Nucl.\ Phys.\ B {\bf 614}, 195 (2001)
[arXiv:hep-th/0107140].
}

\lref\calibr{S.~Gukov, ``Solitons, Superpotentials and Calibrations",
Nucl.\ Phys.\ {\bf B574} (2000) 169,
[arXiv:hep-th/9911011].}

\lref\GVW{S.~Gukov, C.~Vafa, and E.~Witten, ``CFT's From
Calabi-Yau Four-folds,'' Nucl.\ Phys.\ {\bf B584} (2000) 69,
[arXiv:hep-th/9906070].}

\lref\JY{A.~Jevicki and T.~Yoneya,
``A Deformed matrix model and the black hole background
in two-dimensional string theory,''
Nucl.\ Phys.\ B {\bf 411}, 64 (1994)
[arXiv:hep-th/9305109].
}

\lref\DR{K.~Demeterfi and J.~P.~Rodrigues,
``States and quantum effects in the collective field theory
of a deformed matrix model,''
Nucl.\ Phys.\ B {\bf 415}, 3 (1994)
[arXiv:hep-th/9306141].
}

\lref\DiKu{P.~Di Francesco and D.~Kutasov,
``World sheet and space-time physics in two-dimensional
(Super)string theory,''
Nucl.\ Phys.\ B {\bf 375}, 119 (1992)
[arXiv:hep-th/9109005].
}

\lref\poc{J.~Polchinski,
``Classical Limit Of (1+1)-Dimensional String Theory,''
Nucl.\ Phys.\ B {\bf 362}, 125 (1991).
}

\lref\DJ{S.~R.~Das and A.~Jevicki,
``String Field Theory And Physical Interpretation Of D = 1 Strings,''
Mod.\ Phys.\ Lett.\ A {\bf 5}, 1639 (1990).
}

\lref\CGHS{C.~G.~Callan, S.~B.~Giddings, J.~A.~Harvey and A.~Strominger,
``Evanescent Black Holes,''
Phys.\ Rev.\ D {\bf 45}, 1005 (1992)
[arXiv:hep-th/9111056].
}

\lref\Kallosh{
E.~Bergshoeff, R.~Kallosh, T.~Ortin, D.~Roest and A.~Van Proeyen,
``New formulations of D = 10 supersymmetry and D8 - O8 domain walls,''
Class.\ Quant.\ Grav.\  {\bf 18}, 3359 (2001),
[arXiv:hep-th/0103233].}

\lref\wittenbh{E.~Witten,
``On string theory and black holes,''
Phys.\ Rev.\ D {\bf 44}, 314 (1991).
}

\lref\poreview{J.~Polchinski,
``What is string theory?,''
[arXiv:hep-th/9411028].
}

\lref\ponon{
J.~Polchinski,
``Combinatorics Of Boundaries In String Theory,''
Phys.\ Rev.\ D {\bf 50}, 6041 (1994)
[arXiv:hep-th/9407031].
}

\lref\pononc{
J.~Polchinski,
``On the nonperturbative consistency of d = 2 string theory,''
Phys.\ Rev.\ Lett.\  {\bf 74}, 638 (1995)
[arXiv:hep-th/9409168].
}

\lref\pononp{
M.~Natsuume and J.~Polchinski,
``Gravitational Scattering In The C = 1 Matrix Model,''
Nucl.\ Phys.\ B {\bf 424}, 137 (1994)
[arXiv:hep-th/9402156].
}

\lref\hull{
C.~M.~Hull,
``Timelike T-duality, de Sitter space, 
large N gauge theories and  topological field theory,''
JHEP {\bf 9807}, 021 (1998)
[arXiv:hep-th/9806146].
}

\lref\LLM{N.~Lambert, H.~Liu and J.~Maldacena,
``Closed strings from decaying D-branes,''
[arXiv:hep-th/0303139].}

\lref\Strominger{A.~Strominger,
``Open string creation by S-branes,''
[arXiv:hep-th/0209090].}

\lref\Xi{X.~Yin,
``Matrix Models, Integrable Structures, and T-duality of Type 0 String
Theory,'' [arXiv:hep-th/0312236].}

\lref\massiveIIA{
L.~J.~Romans,
``Massive N=2a Supergravity In Ten-Dimensions,''
Phys.\ Lett.\ B {\bf 169}, 374 (1986).
}

\lref\DJ{
S.~R.~Das and A.~Jevicki,
``String Field Theory And Physical Interpretation Of D = 1 Strings,''
Mod.\ Phys.\ Lett.\ A {\bf 5}, 1639 (1990).
}

\lref\AKK{
S.~Y.~Alexandrov, V.~A.~Kazakov and I.~K.~Kostov,
``Time-dependent backgrounds of 2D string theory,''
Nucl.\ Phys.\ B {\bf 640}, 119 (2002)
[arXiv:hep-th/0205079].
}

\lref\GM{
P.~Ginsparg and G.~W.~Moore,
``Lectures On 2-D Gravity And 2-D String Theory,''
[arXiv:hep-th/9304011].
}

\lref\matk{
D.~J.~Gross and N.~Miljkovic,
``A Nonperturbative Solution Of D = 1 String Theory,''
Phys.\ Lett.\ B {\bf 238}, 217 (1990);

E.~Brezin, V.~A.~Kazakov and A.~B.~Zamolodchikov,
``Scaling Violation In A Field Theory Of Closed Strings 
In One Physical Dimension,''
Nucl.\ Phys.\ B {\bf 338}, 673 (1990);

P.~Ginsparg and J.~Zinn-Justin,
``2-D Gravity + 1-D Matter,''
Phys.\ Lett.\ B {\bf 240}, 333 (1990).
}

\lref\KMSZ{
I.~R.~Klebanov, J.~Maldacena and N.~Seiberg,
``Unitary and complex matrix models as 1-d type 0 strings,''
[arXiv:hep-th/0309168].
}

\lref\GW{
D.~Gross and J.~Walcher
``Non-perturbative RR Potentials in the c=1 Matrix Model,''
[arXiv:hep-th/0312021].
}

\lref\FH{
T.~Fukuda and K.~Hosomichi,
``Super Liouville theory with boundary,''
Nucl.\ Phys.\ B {\bf 635}, 215 (2002)
[arXiv:hep-th/0202032].
}

\lref\ARS{
C.~Ahn, C.~Rim and M.~Stanishkov,
``Exact one-point function of N = 1 super-Liouville theory with boundary,''
Nucl.\ Phys.\ B {\bf 636}, 497 (2002)
[arXiv:hep-th/0202043].
}

\lref\Malda{J.~M.~Maldacena,
``The large N limit of superconformal field theories and supergravity,''
Adv.\ Theor.\ Math.\ Phys.\  {\bf 2}, 231 (1998)
[Int.\ J.\ Theor.\ Phys.\  {\bf 38}, 1113 (1999)]
[arXiv:hep-th/9711200].
}

\lref\SW {L.~Susskind and E.~Witten,
``The holographic bound in anti-de Sitter space,''
[arXiv:hep-th/9805114].
}

\lref\Pocr{
J.~Polchinski,
``Critical Behavior Of Random Surfaces In One-Dimension,''
Nucl.\ Phys.\ B {\bf 346}, 253 (1990).
}

\lref\GKP{
S.~S.~Gubser, I.~R.~Klebanov and A.~M.~Polyakov,
``Gauge theory correlators from non-critical string theory,''
Phys.\ Lett.\ B {\bf 428}, 105 (1998)
[arXiv:hep-th/9802109].
}

\lref\Dabh{
U.~H.~Danielsson,
``A matrix model black hole: Act II,''
arXiv:hep-th/0312203.
}

\lref\Da{
U.~H.~Danielsson,
``A Matrix model black hole,''
Nucl.\ Phys.\ B {\bf 410}, 395 (1993)
[arXiv:hep-th/9306063].
}


\baselineskip 18pt plus 2pt minus 2pt

\Title{\vbox{\baselineskip12pt
\hbox{hep-th/0312208}
\hbox{HUTP-03/A088}
\hbox{ITEP-TH-93/03}
  }}
{\vbox{\centerline{Flux Backgrounds in 2D String Theory}
}}
\centerline{Sergei Gukov,$^a$\footnote{$^1$}{On leave from
the Institute of Theoretical and Experimental Physics
and the L.~D.~Landau Institute for Theoretical Physics} 
Tadashi Takayanagi,$^a$ 
and Nicolaos Toumbas$^b$}

\bigskip\centerline{\it $^a$ Jefferson Physical Laboratory}
\centerline{\it Harvard University}
\centerline{\it Cambridge, MA 02138, USA}

\medskip

\medskip\centerline{\it $^b$ Department of Physics}
\centerline{\it University of Cyprus}
\centerline{\it Nicosia, CY-1678, Cyprus}

\vskip .3in \centerline{\bf Abstract}

We study RR flux backgrounds in 
two dimensional type 0 string theories.
In particular, we study the relation
between the 0A matrix model and the extremal black hole in two dimensions.
Using T-duality we find a dual flux background in
type 0B theory and propose its matrix model description.
When the Fermi level $\mu$ is set to zero
this system remains weakly coupled and exhibits a larger symmetry
related to the structure of flux vacua. 
Finally, we construct a two dimensional type IIB background    
as an orbifold of the 0B background.

\noblackbox

\Date{December 2003}

\writetoc

\newsec{Introduction and Summary}

Two dimensional string theory provides a simple and tractable system,
where perturbative string dynamics can be studied exactly using the
connection with the $c=1$ matrix model \matk
(for reviews see e.g. \refs{\KlR,\GM,\poreview}).
Motivated by the relation between the $c=1$ matrix model
and the dynamics of unstable D0-branes \refs{\MV,\KMS,\MTV}, 
recently \refs{\TT,\six} proposed a non-perturbative matrix model
description for type 0 string theory in two dimensions.
According to \refs{\TT,\six}, the type 0B matrix model is defined
as a double Fermi sea version of the usual $c=1$ matrix model,
whereas the 0A matrix model is defined
as a theory of complex random matrices, which describe
open string degrees of freedom in the $D0-\ov{D0}$ system.
These models do not suffer from non-perturbative instabilities
encountered in the bosonic string theory \refs{\ponon,\pononc,\poreview},
and lead to consistent unitary theories as shown by 
\five\ by applying the results of \MoPlRa \ 
%
%
%
\nref\DV{R.~Dijkgraaf and C.~Vafa,
``N = 1 supersymmetry, deconstruction, and bosonic gauge theories,''
[arXiv:hep-th/0302011].
}
\nref\Ma{E.~J.~Martinec,
``The annular report on non-critical string theory,''
[arXiv:hep-th/0305148].
}
\nref\AKKN{
S.~Y.~Alexandrov, V.~A.~Kazakov and D.~Kutasov,
``Non-perturbative effects in matrix models and D-branes,''
JHEP {\bf 0309}, 057 (2003)
[arXiv:hep-th/0306177].
}
\nref\GIR{
D.~Gaiotto, N.~Itzhaki and L.~Rastelli,
``On the BCFT description of holes in the c = 1 matrix model,''
Phys.\ Lett.\ B {\bf 575}, 111 (2003)
[arXiv:hep-th/0307221].
}
\nref\GK{
M.~Gutperle and P.~Kraus,
``D-brane dynamics in the c = 1 matrix model,''
[arXiv:hep-th/0308047].
}
\nref\Sen{
A.~Sen,
``Open-closed duality: Lessons from matrix model,''
[arXiv:hep-th/0308068].
}
\nref\MMV{
J.~McGreevy, S.~Murthy and H.~Verlinde,
``Two-dimensional superstrings and the supersymmetric matrix model,''
[arXiv:hep-th/0308105].
}
\nref\Ka{A.~Kapustin,
``Noncritical superstrings in a Ramond-Ramond background,''
[arXiv:hep-th/0308119].
}
\nref\Tes{
J.~Teschner,
``On boundary perturbations in Liouville theory 
and brane dynamics in noncritical string theories,''
[arXiv:hep-th/0308140].
}
\nref\GKPS{
A.~Giveon, A.~Konechny, A.~Pakman and A.~Sever,
``Type 0 strings in a 2-d black hole,''
JHEP {\bf 0310}, 025 (2003)
[arXiv:hep-th/0309056].
}
\nref\Nak{
Y.~Nakayama,
``Tadpole cancellation in unoriented Liouville theory,''
JHEP {\bf 0311}, 017 (2003)
[arXiv:hep-th/0309063].
}
\nref\KaSt{
J.~L.~Karczmarek and A.~Strominger,
``Matrix cosmology,''
[arXiv:hep-th/0309138].
}
\nref\GKA{
J.~Gomis and A.~Kapustin,
``Two-dimensional unoriented strings and matrix models,''
[arXiv:hep-th/0310195].
}
\nref\AFKS{
O.~Aharony, B.~Fiol, D.~Kutasov and D.~A.~Sahakyan,
``Little string theory and heterotic/type II duality,''
[arXiv:hep-th/0310197].
}
\nref\Muk{
S.~Mukhi,
``Topological matrix models, Liouville matrix 
model and c = 1 string theory,''
[arXiv:hep-th/0310287].
}
\nref\BeHi{
O.~Bergman and S.~Hirano,
``The cap in the hat: Unoriented 2D strings and 
matrix(-vector) models,''
[arXiv:hep-th/0311068].
}
\nref\Jo{
C.~V.~Johnson,
``Non-perturbative string equations for type 0A,''
[arXiv:hep-th/0311129].
}
\nref\ES{
T.~Eguchi and Y.~Sugawara,
``Modular bootstrap for boundary N = 2 Liouville theory,''
[arXiv:hep-th/0311141].
}
\nref\DaDa{
S.~Dasgupta and T.~Dasgupta,
``Nonsinglet sector of c = 1 matrix model and 2D black hole,''
[arXiv:hep-th/0311177].
}
\nref\ADKMV{M.~Aganagic, R.~Dijkgraaf, A.~Klemm, M.~Marino, C.~Vafa,
``Topological Strings and Integrable Hierarchies,''
[arXiv:hep-th/0312085].
}
\nref\DSVY{
J.~de Boer, A. ~Sinkovics, E. ~Verlinde, J.T. ~Yee,
``String Interactions in c=1 Matrix Model,''
[arXiv:hep-th/0312135].
}
\nref\ThA{
D.~M.~Thompson,
``AdS Solutions of 2D Type 0A,''
arXiv:hep-th/0312156.
}
\nref\Janik{J.~Ambjorn, R.~A.~Janik, ``The decay of quantum D-branes,''
[arXiv:hep-th/0312163].
}
\nref\Shih{N.~Seiberg, D.~Shih, ``Branes, Rings and 
Matrix Models in Minimal (Super)string Theory,'' [arXiv:hep-th/0312170].
}
\nref\MW{G.~Mandal, S.~R.~Wadia, ``Rolling 
tachyon solution of two-dimensional string theory,'' 
[arXiv:hep-th/0312192].
}
\nref\StA{
A.~Strominger,
``A Matrix Model for $AdS_2$,''
arXiv:hep-th/0312194.
}
$\!\!\!\!\!\!\!\!\!\!\!\!\!\!\!\!\!\!\!\!\!\!\!\!\!\!\!\!
\!\!\!\!\!\!\!\!\!\!\!\!\!\!\!$
(see also e.g. \refs{\DV - \StA}
for recent discussions of $c=1$ matrix model 
and two dimensional string
theory).

In addition to non-perturbative stability,
an advantage of these models compared to the bosonic string theory
is that they admit backgrounds with RR-fluxes.
In the 0B theory, such backgrounds can be obtained by considering
asymmetric Fermi levels.
On the other hand, in the 0A theory the RR flux is determined by
the parameter $q$, which enters the definition of the dual matrix model
and can be interpreted as the net D0-brane charge \six.
It was pointed out in \refs{\Ka,\five}(see also \six),
that after projecting onto the singlet sector,
the 0A model reduces to the Jevicki-Yoneya model \JY,
which was originally proposed as a candidate for
describing the two dimensional black hole \wittenbh.
More recently, a different matrix model dual
of the black hole was proposed in \KKK.

Motivated by these developments, in this paper we investigate
0A and 0B flux backgrounds from the viewpoint of both 
the matrix model and the two-dimensional space-time gravity.
The 0A matrix model with non-zero value of the parameter $q$
can be weakly coupled (and non-perturbatively stable)
even when the Fermi level $\mu$ is zero.
We argue that in the supergravity approximation,
its ground state can be identified with 
the extremal black hole \BGV\ found in the context
of flux compactifications of type II string theory
down to two dimensions \refs{\GVW,\calibr}.
In particular, we identify the energy of the 0A matrix model
with the ADM mass of the extremal black hole and comment on
the UV/IR relation in open/closed string duality.

It is also useful to consider T-dual backgrounds in type 0B theory
with non zero RR flux.
By analyzing the partition functions of the 0A and 0B matrix models
we demonstrate their relation under T-duality.
We propose a Fermi sea description of a
type 0B background with $\mu=0$ and large lightlike RR flux.
It is curious to note, that it has similar perturbative 
structure with the usual 0B background with zero flux 
and non-zero tachyon condensate. 
Using the results of \CGHS, we also construct a variety
of flux solutions in the effective space-time gravity theory, 
including the solution T-dual to the extremal black hole
and various time-dependent backgrounds with flux.
We provide their Fermi sea description in the matrix model.

Interestingly, the 0B matrix model configuration we analyze
in detail is invariant under a $\Z_2$ symmetry
which interchanges the fermions and the holes.
This symmetry is equivalent to GSO projection on the world-sheet.
Thus orbifolding by this symmetry,
gives a two-dimensional type IIB string theory background. In conclusion
we have a $c=1$ 
matrix model description of two-dimensional type IIB theory. 

The paper is organized as follows.
In section 2 we discuss the relation between the 0A matrix model and
the extremal black hole solution of \BGV. 
In section 3 we study the 0B theory with
non-trivial large RR flux and propose a dual Fermi sea description.
An $SU(2)$ symmetry which mixes particles and holes
enters the description in an interesting way.
In section 4 we consider classical solutions in 0B space-time gravity.
We find a static solution which describes the flux background
as well as time dependent solutions which correspond to 
time dependent Fermi seas. In section 5 we describe a matrix model
dual of a type IIB background as an orbifold of the 0B matrix model.

\medskip\noindent
{\bf Note added:}
While we were completing this manuscript, we noticed 
the paper \GW , which has partial overlap with section 3.
After submitting this paper we received \Dabh\ which also
discusses the relation of the 0A model to the extremal black hole.


\newsec{Flux Condensation in 0A Theory and Black Holes}

\subsec{The 0A Matrix Model}

The 0A matrix model in two dimensions \refs{\six, \TT}
is defined as a sum over complex rectangular random matrices.
The model is expected to describe
the dynamics of $N+q$ D0-branes and $N$ anti D0-branes in the theory.
More precisely, in type 0A theory
we have two types of D0-branes, electric and magnetic.
Each type of D0-branes couple to a RR 1-form potential\foot{The electric 
brane couples to the symmetric combination of the RR 1-form potentials from
the RR sector $(R+,R-)\oplus (R-,R+)$ of the theory,
while the magnetic brane to the antisymmetric combination.}.
In the presence of a Liouville potential in the worldsheet theory,
or a tachyon background $T=\mu e^{\phi}$ in spacetime, only one 
type of brane
is physical \refs{\FuHo,\ARS,\TT,\six}. 
The type of brane allowed depends on the sign of
$\mu$. For positive $\mu$, we have electric branes while for 
negative $\mu$ magnetic branes.

The field content of the model consists of 
an $(N+q)\times N$ complex matrix $\Phi$, corresponding to the open
string tachyon field, and two non-dynamical gauge fields $A^{(1,2)}_0$ 
in the adjoint of $U(N+q)$ and $U(N)$ respectively.
The action is defined to be $S=\beta\int dt L$ with the
Lagrangian
\eqn\laga{L={1 \over 2}\Tr[(D_t \Phi)^{\dag} D_t \Phi-U(|\Phi|^2)],}
where $D_t\Phi=\partial_t\Phi-iA^{(1)}_0\Phi+i\Phi A^{(2)}_0$
is the covariant derivative for the gauge group $U(N+q)\times U(N)$; and
$U(|\Phi|^2)=-\f{1}{2\al}|\Phi|^2+\ddd$
is the tachyon potential.
In \six\ it is
proposed that the integer $q$ corresponds to the background RR flux in
0A theory.

After integrating out the gauge fields and taking the double scaling limit,
the system effectively
reduces to $N$ decoupled non-relativistic fermions moving on a plane,
each with angular momentum $q$. The non-zero value of 
the angular momentum has the effect of 
pushing the particles away from the origin. 
The effective dynamics of the radial motion is then described by the
Jevicki-Yoneya model \JY\ as shown in \refs{\Ka, \five}. This model is
defined by the Hermitian matrix model with the 
following potential (we always set $\al=\f{1}{2}$ on matrix model side)
\eqn\jy{V(x)=-x^2+\f{M}{x^2},} where the deformation parameter
$M$ is related\foot{This relation can be easily found
using the second wave equation (10.8) in \six. After redefining a
new wave function $\psi=\lambda_i^{-\f{1}{2}}\chi$, we get
$-(\f{d^2}{d\lambda_i^2}
+{\lambda_i^2}-\f{q^2-1/4}{\lambda_i^2})\psi=E\psi.$
}
to the quantized flux $q$ via $M=q^2-1/4$. The classical trajectory with
the energy $-\mu$
is given by
\eqn\tra{x(\tau)^2=\mu+\s{M+\mu^2}\cosh(2\tau).}
This model is exactly solvable, so that one can compute
non-perturbative scattering amplitudes explicitly 
as in \refs{\Kl,\Ka,\five} (see also \refs{\Da, \DR}).

It is also possible to compute the free energy in this model.
One can see that 
the density of states $\rho(\ep)$ is given by \refs{\Kl, \Da, \DR}
\eqn\dst{\rho(\ep)=-\f{1}{4\pi}[\psi(\f{1}{2}-\f{i\ep}{2}+\f{q}{2})
+\psi(\f{1}{2}+\f{i\ep}{2}+\f{q}{2})]+ {\rm const},}
where
$\psi(z)\equiv \f{d\log\Gamma(z)}{dz}$. When the flux is large,
$q>>1$, we can expand the expression \dst\ as follows\foot{
To be exact a cut off $L$ of the coordinate $x$ is included in this 
expression as the extra term $\f{1}{\pi}\log L$.} 
\eqn\explo{\rho(\ep)\sim -\f{1}{4\pi}\log
(q^2+\ep^2)-\f{1}{12\pi}\f{q^2-\ep^2}{(q^2+\ep^2)^2}+\ddd.}
Then we can compute the free energy
at Fermi level $-\mu$ \foot{The Fermi level 
is measured from the top of the harmonic 
oscillator potential at $x=0$.}. In particular, when $|\mu|<<q$, we get
(see \DR \ for earlier discussions)
\eqn\free{E=\int^{-\mu}_{-\mu_0} \ep\rho(\ep)d\ep\sim
-\f{1}{8\pi}q^2\log q^2 +\f{1}{24\pi}\log q^2+\ddd,}
where the dots denote analytic terms
in $q$.
The important point is that even if we take $\mu$ to zero,
the perturbative expression is still non-singular due to
 non-zero value of $q$, as can be seen directly from eq. \free.
At the point $\mu\sim q\sim 0$,
the corresponding background of 0A theory describes the linear dilaton
background, which is obviously singular.
We conclude that there are two ways to stabilize
the background: one is to add 
a Liouville potential in the worldsheet theory $\mu e^{\phi}$,
condensing the closed string
tachyon field $T$, and another is to add large background flux $q$.
In this paper we shall consider the latter case.


\subsec{Collective Field Theory}

In order to gain some insight into 
the dynamics of the model, it is useful 
to review briefly the collective field 
theory \refs{\DJ,\poc,\poreview} of the Fermi liquid following \DR.
In the classical limit, the collective 
motions of the liquid can be described 
in terms of a time dependent Fermi surface. 
For small perturbations, the Fermi surface 
consists of an upper and lower part, $p_{\pm}(x, t)$, 
subject to the equation of motion
\eqn\emfs{\partial_tp_{\pm}(x,t)
=x+ {M\over x^3}-p_{\pm}(x,t)\partial_xp_{\pm}(x,t).}
The static solution is given by
\eqn\static{p_{\pm}(x)_{static}=\pm \pi \phi_0(x)=\pm 
\sqrt{x^2-{M \over x^2}-2\mu},} with $-\mu$ being the 
Fermi level. The Hamiltonian can be obtained in terms 
of the fields $p_{\pm}$ by integrating the single particle 
Hamiltonian
$h(p, x)={1 \over 2}p^2 + {1 \over 2} V(x)+\mu$ 
over the Fermi sea
\eqn\Hamil{\eqalign{H&={1\over 2\pi}
\int dx\int_{p_{-}}^{p_{+}}dp h(p, x)
\cr &={1 \over 2\pi}\int dx \left[ {1 \over 6}
(p_{+}^3-p_{-}^3)-{1 \over 2}(x^2 -
{M \over x^2}-2\mu)(p_{+}-p_{-})\right]. \cr}}
From the equation of motion and the Hamiltonian, 
we can deduce the Poisson brackets

\eqn\brackets{\eqalign{
& \{p_{\pm}(x), p_{\pm}(y)\}=-2\pi \partial_x\delta(x-y), \cr
& \{p_{+}(x), p_{-}(y)\}=0. }}

We now change variables from $x$ to the 
variable $\tau$ defined \foot{The turning point of the 
classical trajectory eq. \tra \ occurs at 
$x^2=\mu +\sqrt{M+\mu^2}$ when $\tau=0$, while 
as $\tau\rightarrow -\infty$ we have $x\rightarrow \infty$.} by
\eqn\time {d\tau=-{dx \over \pi\phi_0(x)},}
and expand the fields $p_{\pm}$ about the static background
\eqn\shifted {p_{\pm}=\pm\pi\phi_0\pm{1 \over \pi\phi_0}
\epsilon_{\pm}(\tau, t).} In terms of these, we define 
a scalar field $\ov {S}(\tau, t)$
\eqn\scalar{(2\pi)^{-1/2}\epsilon_{\pm}(\tau, t)=\pm 
\ov{{{\Pi}}}_{{S}}(\tau,t)-\partial_{\tau}\ov{S}(\tau,t),}
to obtain
\eqn\newHam {H={1 \over 2}\int_{-\infty}^0 d\tau 
\left[ {\ov{\Pi}_S}^2 + (\partial_{\tau}\ov{S})^2 
+ {e^{2\tau} \over \sqrt{M +\mu^2}}O({\ov{S}}^3)\right] + E_{0},}
where
\eqn\clasenergy {E_0=-{1 \over 3\pi}\int dx(x^2 - M/x^2-\mu)^{3/2},}
is the classical energy of the static background \DR.
Thus, in the large $\tau$ region, we end up with a relativistic massless
scalar field. As we approach the endpoint 
of the eigenvalue distribution at $\tau=0$, 
non-relativistic self interactions of the scalar 
field $\ov{S}(\tau,t)$ become important.
At the linearized level, the fields $\epsilon_{\pm}$ 
are simply the right and left moving modes 
satisfying $(\partial_t \pm \partial_{\tau})\epsilon_{\pm}(\tau, t)=0$. 
These modes are not independent but are mixed by Dirichlet 
boundary conditions at the endpoint of the eigenvalue 
distribution\foot{We refer the reader in \DR \ for 
how to deal with technical difficulties arising from the 
boundary conditions at the endpoint of the distribution.
 As noted in \DR, for the purposes of perturbation theory, 
one can keep only the right moving field, treating it as an 
independent field, and extend the range of the $\tau$
integration over the whole real line. One identifies 
$\epsilon_{+}(-\tau)=\epsilon_{-}(\tau)$.}.

The position dependent coupling constant scales as $e^{\phi}$ 
in string theory and $e^{2\tau}/\sqrt{M+\mu^2}$ in the matrix model. 
Thus for large $M$ or small $\mu$, we identify $\tau$ with 
the dimensionless linear dilaton spacetime 
coordinate\foot{The coordinate $\phi$ is related to the spatial 
coordinate $\vp$ through $\phi=\s{2 \over \al}\vp$.} (asymptotically) 
as follows
\eqn\relationphitau {\phi=-{1 \over 2}\log M +2\tau=-\log x^2.}
The dominant interactions occur at the endpoint of the eigenvalue 
distribution near $\tau=0$. At this point, the effective spacetime 
coupling constant is given by
\eqn\effcoup {g_{eff} \sim {1 \over \sqrt{M+\mu^2}}.}
It remains small even at $\mu=0$, when the 
deformation parameter $M$ is large.
Thus in this case, we expect the perturbation theory in the matrix model
to remain valid, suggesting a weakly coupled string theory dual background
with $g_{eff}\sim 1/q$.

When $\mu$ is large, we interpret scattering off the endpoint of the
eigenvalue distribution in the matrix model as string scattering off the
tachyon Liouville wall in spacetime. The Liouville wall prevents the strings
from propagating into the strongly coupled region. Near the wall the strings
interact with one another with effective coupling constant of order $1/\mu$
and they reflect back to the weakly coupled region.
At $\mu=0$,
the tachyon background is turned off and we
interpret the scattering as tachyon scattering off a gravitational potential.
The effective coupling constant is of order $1/q$.
The corresponding background is curved.
Indeed, as we argued earlier, the deformation parameter $M$
in the matrix model is related to the
number of units of RR 2-form flux in spacetime.
The background RR flux curves spacetime.
We will provide evidence
that the relevant background at $\mu=0$ is the extremal black hole solution
of \BGV.
Effectively, the flux cuts off the strongly coupled region, `hiding' it
behind the horizon of the black hole.

The classical energy eq. \clasenergy \ of the static 
background is infinite.
We can regulate it by subtracting the energy of the 
corresponding background
at $M=0$ and imposing a long distance cut-off $L$ on the 
coordinate $x$. Then
at $\mu=0$, we find for large $M$
\eqn\zeroen {E_0 = - {M \over 8\pi}\log{M \over L^4}.}
This is of course the leading order contribution to the free
energy computation eq. \free. Using the relation to the linear dilaton
spacetime coordinate $\phi$, we can express the cut-off dependent
divergent piece as
$ -M\phi/4\pi|_{\phi=-\infty}$. Clearly this cut-off corresponds
to an infrared cut-off from the point of view of spacetime,
regulating the infinite volume of spacetime. We shall see that the finite
piece agrees precisely with the ADM mass of the extremal black hole
background, which is obtained after we subtract the same
divergent piece in the gravity theory.

From the point of 
view of the free energy computation, eq. \free, \ using the
density of states $\rho(\epsilon)$, the cut-off corresponds to a
cut-off of high negative energy modes deep in the Fermi sea
\eqn\cutoffs {\mu_0 \sim L^2.}
In this large $x$ region, the open string tachyon field has attained a
large expectation value.
In this sense,
the cut-off corresponds to an ultraviolet cut-off in the open string theory
on the
D-branes. This is a manifestation of a UV/IR relation between the open and
closed string sides of the duality similar to the 
UV/IR relation\foot{See \refs{\Pocr, \GKP}  
for a similar relation in the context of
two dimensional bosonic string theory.} 
in the AdS/CFT correspondence \refs{\Malda, \SW}. 
A probe fermion eigenvalue following the classical 
Fermi trajectory corresponds to a D0 brane/anti-D0 
brane pair decaying into closed string radiation \refs{\MV,\KMS}. 
In the large $x$ region, the fermion is relativistic 
and can be bosonized. It describes a coherent state of 
closed strings. In the open string channel, this state involves 
high frequency off-shell open strings \refs{\Strominger,\LLM}.

An important result from the old studies of this matrix model 
is the fact that
scattering amplitudes involving an odd number of scalars $\ov{S}$ 
vanishes \refs{\JY,\DR,\Kl}. In the perturbative regime, this follows 
from the fact that the cubic interaction 
vertex at $\mu=0$ vanishes on shell. This suggests that there 
should be a field redefinition under which the S-matrix remains 
invariant and makes a $\Z_2$ symmetry $\ov{S}\leftrightarrow -\ov{S}$ 
in the effective action manifest. From the point of view of 
spacetime physics, the collective scalar field $\ov{S}$ describing 
the fluctuations of the Fermi sea is related to the closed 
string tachyon field $T$. The 
spacetime effective action for the tachyon field has such 
a symmetry. Under this symmetry the action remains 
invariant if one also interchanges the electric and magnetic 
RR one-form potentials. From the point of view of the worldsheet 
theory, this operation is $(-1)^{F_L}$ where $F_L$ is 
the worldsheet fermion number:
\eqn\ztwofl{
(-1)^{F_L} : \quad\quad
\eqalign{
T & \longleftrightarrow -T \cr
F^{(e)} & \longleftrightarrow  F^{(m)}
}}

In the presence of a tachyon background $T=\mu e^{\phi}$ this symmetry is
broken spontaneously. Depending on the sign of 
$\mu$ only electric or magnetic branes are allowed 
physical states and so only one type of flux 
can be induced. For large positive $\mu$, the dual matrix model
describes the dynamics of $N+q$ D0 and $N$ anti-D0 
electric branes \foot{At large negative $\mu$, 
we have a dual description in terms of magnetic branes.}. 
At $\mu=0$, we can add $q$ magnetic branes. 
{}From the spacetime point of view, there are 
$q$ branes of each type localized at $\phi=\infty$ 
because of the linear dilaton background. 
Their effect is to induce $q$ units of electric 
and $q$ units of magnetic 2-form flux in spacetime 
\foot{The flux background is stable against nucleation 
of brane anti-brane pairs because of the linear dilaton. 
The pairs cannot be separated and are confined within the 
horizon region of the extremal black hole.}. 
The corresponding black hole has two types of equal 
charges, and the effective action for the closed 
string tachyon is invariant under the symmetry \ztwofl.

Finally, let us comment on the relation between 
the collective scalar field $\ov{S}$ and the spacetime 
tachyon field $T$. Their relation is non-local; 
in momentum space, they are related by a momentum dependent phase, 
the leg-factor.
To find the leg factor, we follow
the method presented in \refs{\KlR,\Kl}.
Using the operator proposed in \TT\
\eqn\leg{\lim_{l\to 0}\int dt e^{iPt} \Tr
e^{-l\bar{\Phi}\Phi},}
we get the following leg factor phase
\eqn\legg{e^{i\delta(P)}=(q^2+\mu^2)^{-i\f{\s{\al}}{2\s{2}}P}
\f{\Gamma(i\s{\f{\al}{2}}P)}{\Gamma(-i\s{\f{\al}{2}}P)}.}
The operator with finite $l$
creates a macroscopic loop with length $l$ 
in the dual Riemann surface. Here we take the limit $l\to 0$ 
to realize a closed string insertion. 
These leg factors have poles at imaginary integer valued 
momenta corresponding to resonances in the euclidean amplitudes 
due to extra discrete states in the theory \KlR. 
These discrete states are remnants of oscillator modes of the string. 


\subsec{The Extremal Black Hole and the Correspondence}

Now let us consider vacuum solutions of 
0A string theory in the presence of RR
flux. The effective low energy action\foot{If we would like to have the
action for the bosonic string, we must rescale
$\al\to \al/2$.} is given by \six
\eqn\actiona{\eqalign{
S_{2d} &=\int d^{2} x \sqrt{- g}
\Biggl[\f{e^{-2\phi}}{2\kappa^2} \Bigl(\f{8}{\al}
+ R + 4(\nabla \phi)^2-a(\nabla T)^2
+\f{2a}{\al}T^2+\ddd \Bigr) \cr & \ \ \ \ \ \
-\f{2\pi\al}{4}(e^{-2T}|F^{(e)}|^2 +e^{2T}|F^{(m)}|^2)
+ q_+ F^{(e)} + q_- F^{(m)} +\ddd
\Biggr], \cr}}
where $a$ is a certain constant.
The familiar linear dilaton vacuum
is given by
\eqn\lds{\phi=\s{\f{2}{\al}}\vp,\ \ \
g_{\mu\nu}=\eta_{\mu\nu}.}
In this background, $q_{\pm}=0$
and the field $\tilde{T}=e^{-\phi}T$ is massless. 
It corresponds to the scalar field $\ov{S}$ 
describing the collective motions of the Fermi liquid. 
The non-singular $\hat{c}=1$ theory is
obtained by condensing the tachyon field $T=\mu~
e^{\s{\f{2}{\al}}\vp}$, which is proposed to be dual to the 0A matrix
model at the Fermi level $-\mu$ and $q=0$  \refs{\six,\TT}.

Since we are interested in solutions
invariant under \ztwofl\
we consider background flux
\eqn\oaflux{ F^{(e)}=F^{(m)} .}
This flux can be induced by $q$ electric and $q$ magnetic 
D0-branes ``at infinity''.
As can be seen from the action \actiona,
there is no tachyon tadpole in this case. 
The tachyon background expectation value can be set to zero.
The background flux generates a potential for the tachyon 
field $V_{f}(T)\sim q^2 \cosh(2T)$. In the small 
$T$ approximation, this is given by
\eqn\pott{ V_{f}(T)\sim q^2 (1 + 2T^2 + \ddd)
=q^2(1 + 2e^{2\phi}\tilde{T}^2+\ddd).} 
The leading quadratic piece corresponds to 
a position dependent mass term for $\tilde{T}$, 
preventing propagation in the strongly coupled region. 
On the matrix model side, this is achieved by the 
deformation potential in \jy.

As in massive type IIA supergravity \massiveIIA,
the presence of RR-flux is equivalent to
a cosmological constant.
Indeed, integrating out $F^{(e)}$ and $F^{(m)}$ in \actiona,
we obtain the following effective action
(here, we set $2\kappa^2=1$ and $a=1$):
\eqn\acta{
S_{2d} = \int d^{2} x \sqrt{- g} \Big[ e^{-2 \phi}
\big( R + 4 (\nabla \phi)^2 + c - (\nabla T)^2
+ {2 \over \a'} T^2 \big) + \Lambda (1 + 2T^2) + \ldots \Big],
}
where $c\equiv 8/\al$ and
\eqn\cosm{\Lambda=-\f{q^2}{2\pi\al}.}
The field equations obtained from the action \acta\ have the form:
\eqn\eomsa{\eqalign{
& R_{\mu\nu} + 2 \nabla_{\mu} \nabla_{\nu} \phi
- \nabla_{\mu}T\nabla_{\nu}T =0 \cr
& R + c - (\nabla T)^2 + {2 \over \a'} T^2
+ 4 \nabla^2 \phi - 4 (\nabla \phi)^2  =0 \cr
& \nabla^2 T  - 2 \nabla \phi \nabla T
+ \big( {2 \over \a'} + 2 \Lambda e^{2\phi} \big) T =0
.}}

A simple class of solutions in this theory can be obtained by
considering a background with zero tachyon field, $T=0$.
In this case, the theory reduces to the dilaton gravity
with negative cosmological constant studied in \BGV.
In particular, classical solutions in such theory correspond
to black holes with mass parameter $m$
(in this coordinate frame, the dilaton is given by $\phi$):
\eqn\sol{ds^2=-l(\phi)dt^2+\f{d\phi^2}{l(\phi)},}
where
\eqn\fl{l(\phi)=\f{c}{4}-e^{2\phi}(\f{\Lambda}{2}\phi+m).} 
These solutions describe Reissner-Nordstrom 
like charged black holes, and the strongly coupled
region is hidden behind a horizon as in \wittenbh.
If we shift $\phi$ by a constant, we can still obtain a solution
shifting $(m,\Lambda)$ appropriately.
In the weakly coupled region at
$\phi\to -\infty$, the solution approaches the linear 
dilaton vacuum \lds \
and the metric is flat.
In this region, the backreaction of the background flux becomes 
negligible. On the matrix model side
this corresponds to the fact that the deformation term in the 
potential \jy\ is
localized near $x=0$ corresponding to the strongly coupling region.

\fig{A plot of the conformal factor $l(\phi)$ for
a non-extremal $(a)$ and extremal $(b)$ black hole.}
{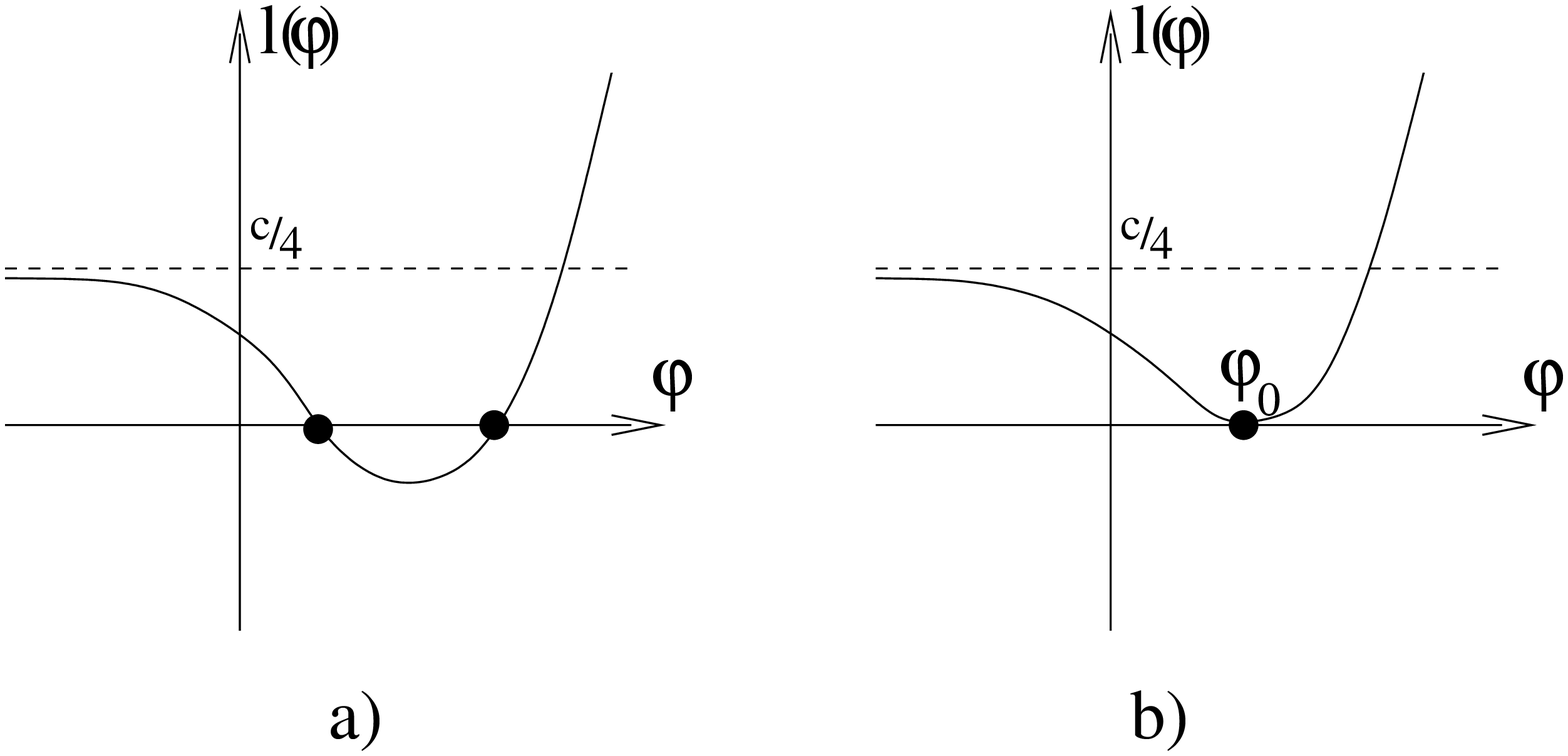}{6.0truein}

The horizon corresponds to the zero of $l(\phi)$. 
When the mass parameter $m$
is bigger than a critical value, $m>m_0$, where
\eqn\massbh{m_{0}=-\f{\Lambda}{4}-\f{\Lambda}{4}\log(-\f{c}{\Lambda}),} 
there are two horizons as in the usual case of Reissner-Nordstrom 
black holes in four dimensions. When $m=m_0$ the inner and outer 
horizons coincide corresponding to the fact that the 
function $l(\phi)$ has a double zero at some point $\phi=\phi_0$,
see Fig.1.
This is the extremal solution of \BGV. The solution \sol \ 
has a naked singularity unless the mass parameter obeys the BPS like 
inequality $m\ge m_0$.

The location of the horizon of the extremal black hole occurs at
\eqn\phihor {\phi_0={1 \over 2}\log(-{c\over \Lambda}),} 
when $l(\phi)=l'(\phi)=0$. Thus the coupling constant near 
the horizon is of order $g_s \sim 1/q$, and the theory remains 
weakly coupled outside the horizon for large $q$. 
This was also the case from the analysis of the collective field 
theory of the Fermi liquid. On the matrix model side, the 
location of the horizon corresponds to the endpoint of the 
eigenvalue distribution. Thus we expect the Fermi 
sea to describe the region outside the horizon of the black 
hole. It is interesting to notice that the extremal black hole metric can be
written in a simple form independent of $\Lambda$
\eqn\rem{l(\ti{\phi})
=\f{c}{4}+\f{c}{2}e^{2\ti{\phi}}(\ti{\phi}-{1 \over 2}),}
by shifting the relation between the dilaton and the spatial coordinate
as
\eqn\shift{\ti{\phi}=\phi-\f{1}{2}\log(-\f{c}{\Lambda}).}
Asymptotically, this shifted coordinate 
corresponds to the variable $\tau$ in
the matrix model.

For the extremal black hole solution the Hawking temperature vanishes. 
The scalar curvature goes to zero as $\phi \to -\infty$ 
and becomes of order one in string units near the horizon. 
In the near horizon limit, the geometry looks like 
two dimensional anti-de Sitter space \BGV. In terms of a 
new space-like variable $u^{-1}=\s{c\over 2}(\phi-\phi_0)$, 
the metric takes the standard form\foot{See \StA\ for a proposal
for a matrix model dual of $AdS_2$ in two dimensional 
type 0A theory. For a further analysis of the classical solution
see also \ThA.}
\eqn\ads {ds^2={1 \over u^2}(-dt^2 + R^2_{AdS}du^2),} 
where $R_{AdS}=\s{2\over c}$ is the radius of 
the anti-de Sitter space. Thus we end up 
with a stringy $AdS_2$. Since the curvature in this region
is of order one in string units, stringy $\al$ corrections 
become important and naively the low energy gravity approximation 
seems to break down. Unlike the case of \wittenbh, \ we do not have 
a full sigma model description of the background geometry.

Let us now obtain the ADM mass of the black hole.
Using the standard formula in 2d gravity we get
\eqn\massr{M_0=\f{1}{2\s{c}}(4m+\Lambda+2\Lambda\phi)|_{\phi=-\infty}.}
For the extremal black hole $m=m_0$ is given by eq. \massbh.
Hence, after subtracting the divergent cut-off dependent 
piece, we obtain for large $q$
\eqn\massexbh{M_0={\Lambda \over 2\s{c}}\log(-{\Lambda \over c})
=-{q^2\over 8\pi\s{2\al}}\log q^2.}
This agrees precisely with the matrix model computation of the 
classical energy
\zeroen, \ where the same cut-off dependent piece is subtracted.
Thus we would like to propose that the 0A matrix model with $\mu=0$
and non-zero $q$ is dual to 0A string theory on the extremal 
black hole background.
More precisely, the matrix model includes stringy $\al$ and 
$g_s$ corrections 
to the low energy gravity analysis. Thus it describes 
a `quantum' black hole. 
We believe that non-extremal black holes
correspond to deformations of the 0A matrix model 
with Wilson lines\foot{Here note that we have two different gauge
fields in the matrix model side: 
diagonal (NS) and relative (R).} as in reference  \KKK\
leading to higher energy configurations.

One important evidence
for this proposal is as follows. As we argued before, on the 
matrix model side, the scattering amplitude of an odd number
of tachyon fields $T$ vanishes \refs{\JY,\DR,\Kl}.
On the gravity side, this just follows from the ${\Z_2}$ symmetry \ztwofl\
of the theory in the symmetric flux background $F^{(e)}=F^{(m)}$.


\subsec{Tachyon dynamics in the Black Hole Background}

Now, let us consider a weak tachyon field in the background
of the extremal black hole \sol\--\fl.
{}From \eomsa\ we find the equation of motion for the massless
tachyon mode $\tilde T = e^{-\phi} T$,
\eqn\tteom{
l^2 \tilde T'' + ll' \tilde T'
+ (ll' - l^2 + l(\f{2}{\a'}+2\Lambda e^{2\phi})) \tilde T
- {\p^2 \over \p t^2} \tilde T =0
}
where the prime denotes the derivative with respect to $\phi$,
which plays the role of the radial coordinate
in the Schwarzschild gauge \sol.
Notice, that the coefficient of the linear derivative term
vanishes if we write \tteom\ in terms of the original
spatial coordinate $x$, such that ${d\phi \over dx} = l(\phi)$,
\eqn\tteomx{
{\p^2 \over \p x^2} \tilde T
+ (ll' - l^2 + l(\f{c}{4}+2\Lambda e^{2\phi})) \tilde T
- {\p^2 \over \p t^2} \tilde T =0.
}

Let us look at the tachyon mode with energy $\omega$
\eqn\tomega{ \tilde T = \tilde T_{\omega} e^{- i \omega t}. }
Substituting this into \tteomx, we find that the tachyon
equation of motion takes the form of the one-dimensional
Schr\"odinger equation
\eqn\schrod{
\left( \f{d^2}{dx^2} + \omega^2 - V_{{\rm eff}} \right) 
\tilde T_{\omega} =0,
}
with the effective potential
\eqn\vefft{
V_{{\rm eff}}
= - l \left( l' - l + \f{c}{4}+2\Lambda e^{2\phi} \right)
.}
Using the explicit form \fl\ of the conformal factor $l(\phi)$
it is easy to see that $V_{{\rm eff}}$ vanishes in
the asymptotic region, $\phi \to - \infty$, because
the expression in brackets in \vefft\ goes to zero in this limit.
Similarly, since $l(\phi_0)=0$ the effective potential
vanishes at the horizon as well.
At other points, $V_{{\rm eff}}$ is a positive function
with a finite maximum of the order of $c^2 \sim (\a')^{-2}$.
Notice that in the supergravity limit\foot{Since we are using the 
effective field theory of two dimensional gravity, we do not trust
the result beyond this low energy approximation.}, $\a' \to 0$,
the height of the barrier in the effective potential
goes to infinity, so that the tunneling from the asymptotic
region, $\phi \to - \infty$, to the near-horizon region,
$\phi \simeq \phi_0$, is highly suppressed. Thus low energy tachyons 
reflect off a barrier.

Taking backreaction into account, 
one would expect the resulting state to be a non-extremal black 
hole whose metric is given by \sol \  with $m > m_0$. In matrix model 
variables, the perturbation in the metric is of the order 
of $\delta m/x^4$, where $\delta m$ is the energy above extremality. 
Let us now consider a pulse on the Fermi sea of 
width $\delta x$ and height $\delta p$ as in \poreview. 
The analysis in \poreview\ (see also \pononp) shows that in order to 
produce a gravitational effect of order one we need 
$(\delta p)^2 \sim x^2$, i.e. the height of the pulse is comparable 
to the height of the whole sea. 
For such a large pulse, the tachyon self 
interaction is then also of order one and we cannot get into a 
situation where the gravitational interaction 
dominates \foot{The fermions are of course free. 
Tachyon interactions arise due to a strong 
dispersion in the pulse \poreview.}. 
So we cannot be sure that it is possible to make a non extremal 
black hole by sending in tachyon pulses from the asymptotic region 
as described in \BGV. Perhaps the reflection amplitude in the matrix model 
can describe such non-extremal black hole states and their 
subsequent Hawking evaporation to the extremal black hole endpoint. 
It would be interesting to understand how in more detail. 
Notice also that the semiclassical gravity analysis breaks down 
in the near horizon region because the curvature is of order the 
string scale. Unfortunately, we do not have a full string sigma model 
to resolve this puzzle. 

Let us now consider the reflection amplitude
(or two point function)
$S(P)$ ($P$ is the momentum in the $\vp$ direction)
 in this background.  In the fluctuation
analysis around this background
we can easily see that it includes
the factor 
\eqn\legp{S(p)\propto q^{-\s{2\al}iP},} 
due to the shift eq. \shift. 
Even though we cannot obtain the exact reflection amplitude
in the background \sol \ and \rem \ due to the break down of 
low energy analysis, we may expect that the pole structure
can be computable in the effective gravity theory. 
We can find an approximate
expression by replacing
$\ti{\phi} e^{2\ti{\phi}}$ with $e^{2\ti{\phi}}$,
which reduces the computation to
the familiar one in the 2D black hole of \wittenbh.
Using the results\foot{Note that
the convention $\al=2$ in bosonic string
is equivalent to $\al=1$ in 0A string. We also have the relation
$r=-2\phi$.} in \DVV, we find 
the momentum dependent factor
\eqn\legv{S(P)\propto
\f{\Gamma(i\s{\f{\al}{2}}P)~\Gamma^2(\f{1}{2}-i\s{2\al}P)}
{\Gamma(-i\s{\f{\al}{2}}P)~\Gamma^2(\f{1}{2}-i\s{\f{\al}{2}}P)}.}

Now let us compare the results from spacetime computation 
\legp \ and \legv\ with
the matrix model result \legg, though we cannot expect the exact
matching between them.
The
factor \legp\ in the spacetime analysis is indeed included in the matrix
model result \legg\ when $\mu=0$. The position\foot{Here we cannot
expect that \legv \ does exactly agree with \legg\ because the former is a
sort of a minisuperspace computation and the latter is an exact computation
on the matrix theory side.} of poles in \legv \ also agrees
with that in \legg. These provide more supports of our proposal.

\subsec{Matrix Model Thermodynamics}

The finite temperature free energy of the 0A matrix model has been 
essentially
computed in \DR. Up to one loop in the $1/q$ expansion, we have
\eqn\Tfree{F=-{1\over 8\pi}q^2\log ({q^2 \over L^4}) 
+ {1 \over 24\pi}[1+(\pi T)^2]\log ({q^2 \over L^4})+ \ddd,}
where $T$ is the temperature
and $L$ is an infrared volume cutoff \DR. 
The temperature dependent piece shown
is a one loop effect. The thermal entropy is given by
\eqn\entropy{S= -\f{\pi}{12}T \log ({q^2  \over L^4})+\ddd.}
This thermal ensemble corresponds to a gas of massless scalars on a line of
size $\sim \log({L^4 \over q^2})$.

From the spacetime point of view, we interpret the thermal ensemble as a 
gas of tachyons in the asymptotic region of the extremal black hole 
background. Typically, such an ensemble would be unstable against 
gravitational collapse: one has a thermal energy density at temperature 
$T$ and in infinite volume the total energy is infinite. However, 
in two dimensional string theory the linear dilaton term in the 
worldsheet action implies that the gravitational coupling vanishes 
exponentially in the asymptotic region and so the backreaction of 
the thermal gas may be
 negligible. Thus we would expect that the higher loop 
contributions to the free energy are well behaved, as it is suggested 
by the matrix model computations of \DR.

Notice also that the Euclidean geometry
of the extremal black hole is smooth and covers the 
region outside the horizon. If we periodically identify 
asymptotic time at $\phi=-\infty$ with period $2\pi \beta$, 
we still obtain a smooth geometry without any conical 
singularity. As we move toward the stringy $AdS_2$ region, 
the effective temperature increases and becomes of order 
the string scale. At this point, the gravity analysis should 
be replaced with either the full string theory or the dual matrix model.

Of course, one would like to understand if non-extremal black
holes are states within the matrix model.
These thermal states have an extra contribution
to the free energy of order $\delta F \sim q^2T$. 
Unlike eq. \Tfree \ this looks like a genus zero 
contribution. Thus at large $q$, we do not expect to see 
a phase transition in the perturbative regime we explored\foot{
We thank S. Minwalla for many useful discussions
on these and related issues.}.
Instead we can speculate that these states can be 
understood as deformations
of the matrix model by winding modes as in \KKK.


\newsec{Type 0B String with RR-flux}

Here we would like to consider the T-dual of the 0A background
with RR-flux $q\neq 0$ discussed in the previous section.
Consider the 0A theory compactified on
a Euclidean time circle of radius $R$.
In this theory one can define two Wilson line operators with 
winding number $n$ by
$\Omega^n_{\pm}=\Tr (e^{in\int A^{(1)}})\pm \Tr
(e^{in\int A^{(2)}})$, where $A^{(1,2)}$
denote the gauge fields on branes and anti-branes respectively.
These Wilson lines correspond to the winding modes in 0A string theory.
After T-duality they become momentum modes, which
describe fluctuations of the two Fermi seas in the 0B model.
Similarly, the Fermi sea fluctuation modes in the original
0A theory are T-dual to the Wilson lines in the 0B model\foot{
The detailed structure of these T-duality transformations
is studied independently in \Xi.}.

We wish to apply the T-duality transformation
to the RR-flux background discussed in the previous section.
First, let us consider T-duality applied to the Euclidean
continuation of the black hole solution \sol.
We obtain the following metric
\eqn\solb{ds^2=\f{d\theta^2+d\phi^2}{l(\phi)},}
where $\theta$ is the T-dual time direction.
Similarly, the string coupling is given by 
\eqn\strc{g_{s}=\f{e^{\phi}}{\s{l(\phi)}}.}
In the region $\phi \simeq \phi_0$,
we still find an $AdS_2$ geometry
except now we approach the boundary of the space.
This region seems to be strongly coupled,
which naively contradicts the fact that
the original 0A description is non-singular.
As we shall see below, gravitational effects prevent
particles from entering the strongly coupled region.
Note, that in the present case the situation is different from
the T-dual description of the usual black hole \wittenbh.
The latter theory is described by the sine-Liouville theory, 
which implies a closed string tachyon condensate.
In our case there is no tachyon condensation,
as can be seen directly from the $T \lr -T$ symmetry
of the background configuration.

Now, let us consider the T-duality action on the RR fields.
Under T-duality, the non-dynamical fields $F^{(e)}_{01}$ and $F^{(m)}_{01}$
in 0A theory transform into spatial components of the `electric'
and `magnetic' 1-form fields $F$ and $*F$,
which are dynamical in the 0B theory
(locally, we can write $F=dC$).
In other words, we conclude that the 0B dual
of the flux background \oaflux\ can be described by
the self-dual (or anit-self-dual) flux configuration
\eqn\selfdual{ F_+ = f \quad , \quad F_- =0. }
where $F_{\pm}$ are the light-like components of the flux $F$.
Such configurations are similar to instanton solutions
in four-dimensional gauge theory.
As we shall see below, the flux configurations \selfdual\ play
a special role in 0B theory as well. In particular, they automatically
satisfy the tachyon tadpole condition.

Before we proceed to describing specific solutions
in 0B theory with RR flux, we point out that
the non-constant modes of the flux $F_{-} \sim f_{n} e^{in(t-\phi)}$ 
correspond to $\Omega_{-}^n$. 
While generic deformations by $\Omega_{-}^n$ may lead to charged
non-extremal black hole states, {\it cf.} \KKK,
here we will be mainly interested in deformations by constant RR-flux.


\subsec{Fermi Sea of Type 0B with RR Flux and $|\mu|>f$}

Next let us consider the 0B Fermi sea dynamics
using the formalism of \refs{\poreview, \poc}.
The ground state of the two Fermi seas (left $x<0$ and right $x>0$) 
in 0B model with $\mu>0$ is defined by 
\eqn\fermisea{p^l_{\pm}=\pm \s{x^2-2\mu},\ \ \ 
p^r_{\pm}=\mp \s{x^2-2\mu}.}
One can parameterize the fluctuations 
of each Fermi sea as follows
\eqn\fermi{p^l_{\pm}=\mp x \pm \f{\ep_\pm^{l}}{x},\ \ \
p^r_{\pm}=\mp x \pm \f{\ep_{\pm}^{r}}{x}.}
If we use the spatial coordinate $\phi$, such that $e^{-\phi}\sim x^2$,
then we can define two collective fields $S_{NS}$ and $S_{R}$ which in
the asymptotic region $|x|>>1$ are given by
\eqn\fieldt{\ep^l_{\pm}+\ep^r_{\pm}\sim 
\pm \de_{t}S_{NS}-\de_{\phi}S_{NS}, \ \  \ \ \
\ep^l_{\pm}-\ep^r_{\pm}\sim 
\pm \de_{t}S_{R}-\de_{\phi}S_{R}.}
The canonical ground state corresponds 
to $S_{NS}=-2\mu \phi$ and $S_{R}=0$, 
which is obtained by adding the Liouville term
$\int \mu\phi e^{\phi}$ in the 0B worldsheet theory. 

Now we can consider 
perturbations by adding RR flux $F$ in spacetime. We consider\foot{
Note that the RR vertex operator has the Liouville dressing
$e^{\phi}$. The RR field strength $F=dC$ is defined by multiplying with
$e^{-\phi}$ so that we obtain a massless field as in the NSNS sector.}
 the case of constant RR field strength
$F_{\phi}=2f$, 
which is equivalent\foot{To be exact 
we need to take into account the leg factor. However in our example that is 
given by a constant and can be absorbed in $f$.}
 to $S_{R}=2f\phi$ ($S_{R}$ corresponds to the RR 
potential $C$). 
Thus this configuration is described by shifting the Fermi levels
on each side of the potential asymmetrically. In the asymptotic 
region we obtain
\eqn\fermib{p^l_{\pm}\sim \mp x \pm \f{\mu-f}{x},\ \ \ \ \ 
p^r_{\pm}\sim \mp x \pm \f{\mu+f}{x}.}
For large positive $\mu$, large enough so that $\mu > |f|$,
the Fermi sea is described in phase space as in Fig.2.
Even though this flux background is 
perturbatively stable, it will eventually decay due to 
non-perturbative effects
(D-instantons) to a state for which the Fermi levels on the two sides are 
equal. When $|\mu|>>|f|$, such tunneling effects can be neglected.
The asymmetric Fermi sea takes a very long time to decay in this case. 
\fig{The Fermi sea for the 0B flux background when $\mu>f>0$}
{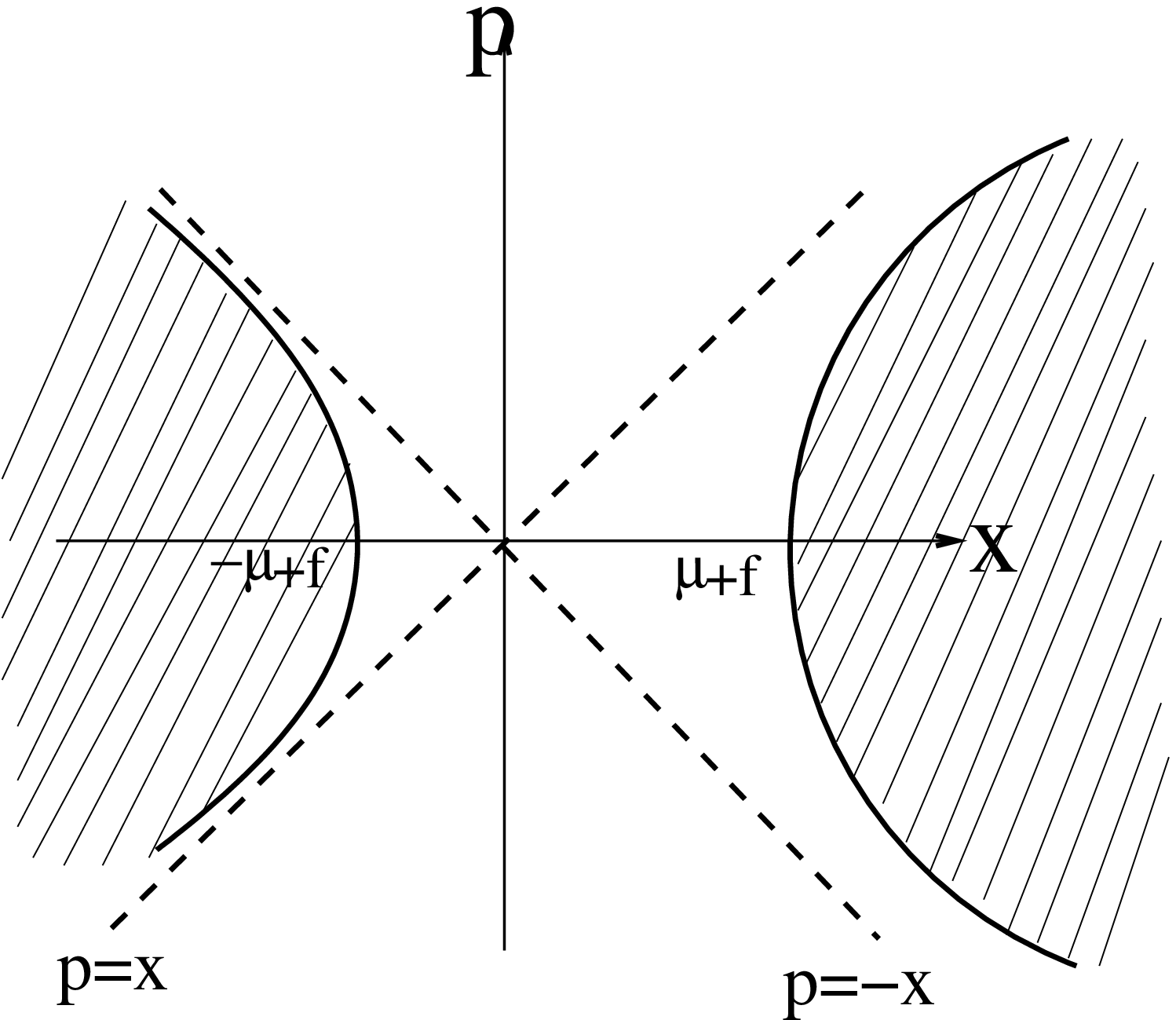}{3.0truein}

When we come to the point $|\mu|=|f|$,
the Fermi sea includes the strongly coupled region $|x| \sim 0$, 
and non-perturbative effects become important.
The situation is similar to the case without flux and $\mu=0$.
One however can go over the barrier at $\mu=0$ by considering
negative $\mu$ and obtain a dual
perturbative
 description setting $T \lr -T$ in spacetime \six.
Thus we may expect that we can go over the barrier
$|\mu|=|f|$ without encountering a phase transition.
As we will argue later, this is indeed the case. For this purpose let us 
examine the partition function (free energy) in the next subsection. 


\subsec{Free Energy and T-duality}

We can also compute the perturbative free energy\foot{
More precisely, this is a Legendre transformation of the usual free
energy with a fixed fermion number.} or 
partition function. It is defined by \refs{\DR, \KKK, \six}
\eqn\defpar{\f{\de^2 \Gamma(\mu)}{\de \mu^2}=\rho(\mu).}
Using this definition, we get the 0A model partition function
(${\cal{Z}}\equiv 2\pi R\Gamma(\mu)$)
\eqn\parta{\eqalign{{\cal{Z}}_A&= 2\Re\Biggl[-R(\mu/2+iq/2)^2
\log (\mu/2+iq/2)-\f{1}{24}
\log(\mu/2+iq/2)(2R+\f{1}{2R}) \cr
&+\sum_{m=1}^{\infty}(\mu/2+iq/2)^{-2m}f_{m}(2R)\Biggr].\cr}}
We have defined the function $f_m$ as 
\eqn\deff{f_{m}(y)=(2m-3)!\ 2^{-2m}\sum_{k=0}^{m}R^{-2k}
\f{|2^{2(m-k)}-2||2^{2k}-2||B_{2(m-k)}||B_{2k}|}{2(m-k)!2k!},} where
$B_{2m}$ are the Bernoulli
numbers ($B_{2m}=(-1)^{m-1}|B_{2m}|$). Notice also the property
\eqn\tdualf{f_{m}(y)=y^{-2m}f_{m}(\f{1}{y}).}
 
On the other hand,
one can also compute the 0B partition function (at radius $\ti{R}$)
with RR flux $f$
by considering the two Fermi sea picture with Fermi levels
$\tilde{\mu}+f$ and $\tilde{\mu}-f$ as follows
\eqn\partb{{\cal{Z}}_B=-\f{\ti{R}}{4}(\ti{\mu}+f)^2
\log (\ti{\mu}+f)^2
-\f{1}{48}\log (\ti{\mu}+f)^2(\ti{R}+\f{1}{\ti{R}})+\sum_{m=1}^{\infty}
(\ti{\mu}+f)^{-2m}f_{m}(\ti{R})\ + (f\lr -f).}
Then, it is easy to see that ${\cal{Z}}_A={\cal{Z}}_B$ 
if we set\foot{A similar result can be also found in 
$c=0$ matrix model \KMSZ.} 
\eqn\relation{\ti{\mu}=\mu R, \ \ \ \  f=iqR,}
and use the T-duality
relation $\ti{R}=\f{1}{2R}$ at $\al=\f{1}{2}$.
The relation \relation\ is very natural from the following observation.
If we take time-like T-duality of type 0A theory, we generally get a 
different theory called `type 0B*' theory from
the usual type 0B theory
in the same way 
as we get type II* theory \hull\ .
The difference between 0B and 0B* appears in the Ramond sector 
and the latter has the wrong sign in front of kinetic term of
RR-fields.
To map the RR-flux in 0B* to 0B we have to multiply it by $i$.
This explains the correspondence described above.


\subsec{Type 0B with RR flux and $|f|\geq |\mu|$}

Let us return to the original problem: the interpretation of the background
with large RR flux $|f|\geq |\mu|$. Let us first consider the  
case $\mu=0$ and $f>0$. Naively the double Fermi 
sea picture looks ill-defined.
Actually there should be no problem in understanding such a background since
the type 0B theory is non-perturbatively well defined \refs{\TT,\six}.
Physical quantities such as scattering amplitudes and the
partition function remain finite. To understand such a
background non-perturbative corrections may turn out to be important.
The best way to analyze this case is to examine
the non-perturbative expressions for the
partition function and amplitudes.
Interestingly even if we set $\mu=0$,
the partition function remains well behaved having a perturbative
expansion with respect to $f^{-1}$ as can be seen from
\partb . One naturally expects
this from the point of view of T-duality from the 0A case,
where the partition function at level $\mu=0$ has
a perturbative expansion in terms of $1/q$.
Scattering amplitudes also have the same property \refs{\five, \MoPlRa}.

Now, if we take the
T-dual\foot{
To be exact we have the 0B* theory as 
T-dual of 0A and furthermore we have to
consider its continuation to 0B theory. As we will discuss later
this is indeed possible.} of the 
0A flux background $\mu=0$ and $q>0$, 
we must add the flux \selfdual\ with only one 
light-like component $F_{+}$ (or equally $F_{t}=F_{\phi}$).
Indeed at $\mu=0$ we can add both components $F_\phi$ and $F_t$ \six.
This is the dual of having both $F^{(e)}=F^{(m)}$ in the 0A model.
In this case we have non-zero expectation values
only for $\ep^{l,r}_{-}$ in \fieldt.
Naively then we get a Fermi sea which does not
correspond to a static background in
the two dimensional gravity theory at low energies. 
The Fermi levels on the left and the right do not
match near $x=0$ and this region should be included
in the occupied phase space.
So one would expect a time dependent configuration to result.  
However, we expect that
the relevant background is time independent as in the 0A dual picture.

It is easy 
to obtain a static configuration by a small modification of the previous
configuration with only $\ep^{l,r}_{-}$.
This is given by the two
Fermi seas $FS1$ and $FS2$ defined by
 (see Fig.3 below)
\eqn\fermisead{FS1=\{(x,p)|x<-\s{p^2+f}\},\ \ \ 
FS2=\{(x,p)|p\geq -x,\ p<\s{x^2+f}\}.}

\fig{The Fermi sea for the 0B flux background when $\mu=0$ and $f>0$}
{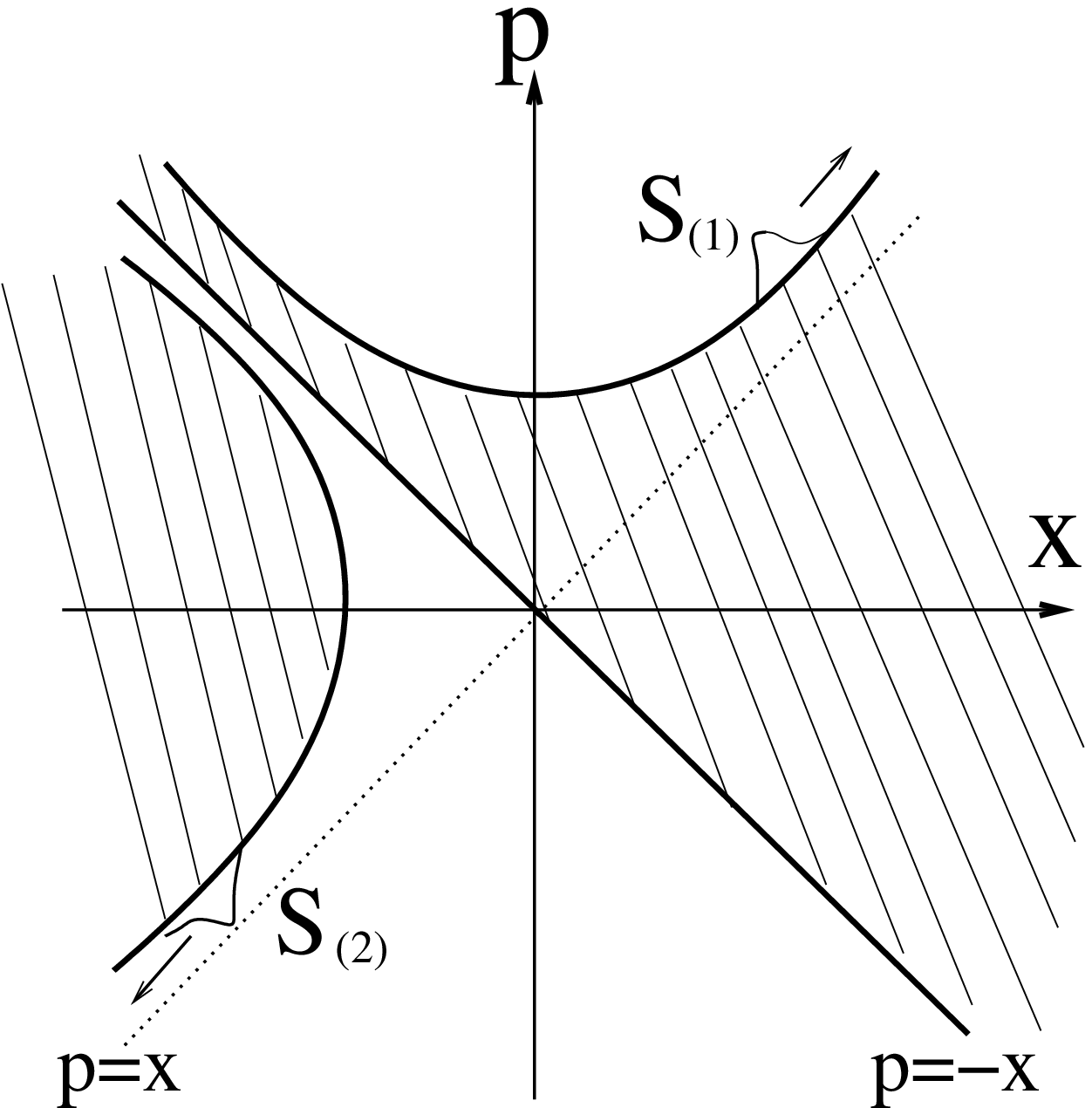}{3.0truein}

The important property of this configuration is that it is invariant under
the following $\Z_2$ transformation\foot{
In the same way we can also define another $\Z_2$ transformation
by $(x,p) \to (p,x)$ and particle-hole exchange. This 
corresponds to another $\Z_2$ action $T\to -T,\  F_{\pm}\to \mp F_{\pm}$.} 
defined in \KlR \six
: transform $(x,p)$ into $(-p,-x)$ and replace
a fermion with a `hole'.
{}From the closed string point of view, this $\Z_2$ transformation
acts as
\eqn\ztwofb{
(-1)^{F_L} : \quad\quad
\eqalign{
T & \longleftrightarrow -T \cr
F_{\pm} & \longleftrightarrow \pm F_{\pm}
}}
The compact RR scalar $C$ is dualized by this action. 

Let us consider a spacetime interpretation of this configuration.
We will provide technical details on this in the next subsection.
It is useful to note that
the axion $C$ is compact at the self-dual radius \six  \ and is 
indeed invariant under the $\Z_2$ symmetry.
Then we can introduce two other scalar fields by
using the $SU(2)_L \times SU(2)_R$ current algebra
\eqn\selfr{\cos(C^{(L,R)})=\de_{\pm} \ti{C}^{(L,R)},
\ \ \ \sin(C^{(L,R)})=\de_{\pm} \ti{\ti{C}}^{(L,R)}.}
Notice that $\ti{C}$ is invariant under the symmetry \ztwofb,
while $\ti{\ti{C}}$ is not. 
Now, from this fact there is an important possibility that
the required modification corresponds
 to adding the flux 
$\ti{F}_{-}=\de_{-}\ti{C}$
, in addition to $F_{+}$. As we will see in the next subsection, 
this can be realized by considering a mixed
state\foot{The terminology `mixed state' refers
to a state involving superposition of particle
and hole creation operators (see section 3.4).}
of particles and holes, which can be approximated
by the Fermi sea \fermisead \ in the large $f$ limit. 
This configuration
represents a type 0B background with RR fluxes $F_{+}=\ti{F}_{-}=f$
and no tachyon. Notice that since it is invariant under the $\Z_2$
symmetry, the tachyon tadpole is zero and we can set $\mu=0$ consistently.
It is also possible to show that 
the corresponding classical solution 
in the effective gravity theory is indeed static as 
we will see in the next section.

On the other hand, if we consider an exact 
pure state represented by \fermisead, \ this 
should be a rather complicated quantum state
of D-branes which cannot be approximated in the 
classical effective gravitational theory. This can be seen from the fact
that with only the flux $F_{+}$ turned on, 
then we cannot construct static classical solutions as is obvious 
from the later analysis in section 4.

The difference between this Fermi sea and the usual 
Fermi sea at level $-\mu$ 
without
flux is that it includes the strongly coupled region near $x \sim 0$.
Naively, this means that in this case non-perturbative effects (D-instantons) 
are important.
However, as we argued before, the perturbative 
region is well decoupled from the strongly coupled region when $f$ is 
large. Thus we may neglect non-perturbative effects for large $f$.
Below we will
discuss the perturbative behavior. 

In the case at hand, we have two collective scalar fields $S_{(1)}$ and 
$S_{(2)}$ describing the fluctuations of the two 
Fermi seas $FS1$ and $FS2$ respectively\foot{The Fermi sea $FS2$
has another Fermi surface at $p=-x$. However, we do not consider
the corresponding collective field because due to tunneling effects
this surface is highly smeared. Part of it is hidden inside the strongly 
coupled region.}. It is useful to define
$S_{(\pm)}=S_{(1)}\pm S_{(2)}$ by taking the 
corresponding linear combinations.
We can identify these with the 
spacetime fields as follows. Since any scattering amplitude 
of an odd number of fields $S_{(-)}$ vanishes in the matrix model,
it must be identified with the tachyon field $T$,
which is also odd under the symmetry \ztwofb.
On the other hand, the field $S_{(+)}$ is invariant
under the $(-1)^{F_L}$ and we can identify it with
the RR scalar $C^{(L)}+\ti{C}^{(R)}$. 
Thus we obtain 
\eqn\rela{S_{(-)}=Te^{-\phi},\ \ \ S_{(+)}=C^{(L)}+\ti{C}^{(R)}.}

Furthermore, we argue that the collective field theory is essentially 
the same as the usual 0B model (with no flux) setting $\mu=f$.
The
perturbative amplitudes are the same as in usual 0B model at non-zero 
$\mu$ and no flux except 
that we must replace $f$ with $\mu$. The partition function is indeed
given by after the same replacement (see \partb). 

It will not be simple to derive 
the perturbative amplitudes on the string theory side because the linear 
dilaton theory is perturbed not only by the RR-vertex operator for $F_{+}$
but also by the operator for $\ti{F}_{-}$. The latter cannot be written
in the RNS formalism in a simple way. The Green-Schwarz formalism 
may be required to work out the amplitudes.
We leave this as a future exercise.
However, we would like to point out that
it would not be so surprising to get the same results 
as the usual $N=1$ Liouville theory. Indeed in the linear dilaton theory 
the free field correlation functions of NSNS and RR-sector vertex 
operators remain essentially the same (up to leg factors) under the exchange
of NSNS ones with RR ones as shown in \DiKu.

Finally we would like to mention that it is straightforward to
generalize this to the case $|f|>|\mu|>0$. It corresponds to the 0B
background with
fluxes $F_{+}=\ti{F}_{-}=f$ and tachyon $T=\mu e^{\phi}$. 
The Fermi surface is given by replacing $f$ with $f+\mu$ for $FS1$
and with $f-\mu$ for $FS2$.

\subsec{Fermi Sea Picture of $SU(2)$ Rotation}

When the tachyon background field is zero $\mu=0$, the axion field $C$ is 
compact at the self-dual radius so that the theory is S-dual 
(electro-magnetic dual) at this point \six. If we neglect non-perturbative
effects, the theory is invariant under the shift symmetry $C\to C+\alpha$ 
\foot{Non-perturbatively, this symmetry is violated by D-instanton 
effects and becomes discrete. Since this discrete shift symmetry is a 
gauge symmetry, the axion field is compact.}.
The left and right moving currents are 
given by $J^{3(L)}=\de_{+}C$ and $J^{3(R)}=\de_{-}C$ (see also \five). 
Actually these are just the components of the RR flux $F_{+}$ and $F_{-}$ 
respectively.
Thus in the asymptotic region, we can define new scalar fields $\ti{C}$ 
and $\ti{\ti{C}}$ by the following $SU(2)$ rotation of the current algebra
\eqn\selfrr{J^{1}_{(L,R)}=\cos(C_{(L,R)})=\de_{\pm} \ti{C}_{(L,R)},
\ \ \ J^{2}_{(L,R)}=\sin(C_{(L,R)})=\de_{\pm} \ti{\ti{C}}_{(L,R)}.}

We would like to discuss the matrix model interpretation of this
SU(2) symmetry. Note that this symmetry is manifest in the asymptotic region.
Since we take the linear dilaton background as the ground state, 
corresponding to setting $\mu=0$ in the matrix model,
we define the creation
operator of a hole $a^{(1,2)\dagger}_i$ and that of a particle (fermion)
$b^{(1,2)\dagger}_i$ as follows (see Fig.4)
\eqn\cran{a^{(1,2)\dagger}_i=\psi^{(1,2)}_{\ep_i}\ \ (\ep_i<0), 
\ \ \ \ \  b^{(1,2)\dagger}_i=\psi^{(1,2)\dagger}_{\ep_i}\ \ (\ep_i>0),}
where $(1,2)$ denotes each of the
two Fermi seas on the left and right. $\ep_i$ is the 
energy of each fermion labeled by $i$, which runs from $1$ to $N$. 
We can define the annihilation
operators in a similar way. Fluctuations of the Fermi 
surface are described by correlated particle/hole pairs.

\fig{The particle and hole creation operators.}
{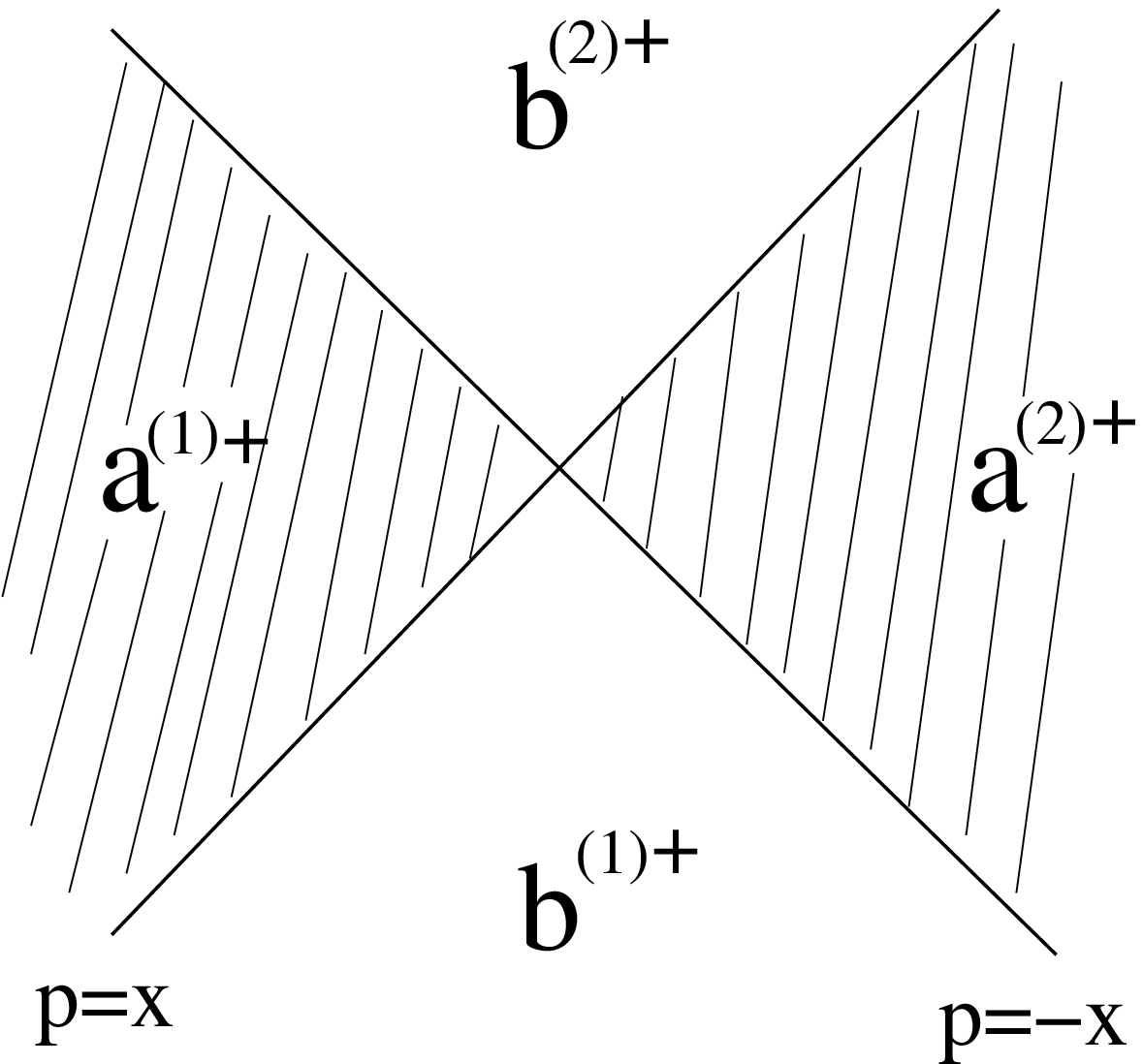}{3.0truein}

Let us consider those 
operations which preserve the total energy. These will turn out to be 
the required symmetries. 
We are interested in only
linear transformations of 
$a^{(1,2)\dagger}_i$ and $b^{(1,2)\dagger}_i$. (Below we  
omit the index $i$ for simplicity.)
Since these 
four complex fermions have the same energy $\ep_i$, we get a $U(4)$ 
symmetry. This is divided into 
$SU(2)_{L}\times SU(2)_{R}$, 
which corresponds to
the decomposition into left and right parts. One $U(1)$ factor
included in $U(4)$ 
amounts to shifting the phase of all fermions on both sides of 
the Fermi sea in the same way. 
It corresponds 
to the shift of tachyon field $T_{L,R}$ 
as can be seen from the bosonization formula 
$\psi^{(1)}=e^{i\f{T+C}{2}},\ \ \psi^{(2)}=e^{i\f{T-C}{2}} $.  
The remaining $SU(2)_{L}\times SU(2)_{R}$ symmetry is identified
with the $SU(2)$ rotation of RR field. 

The $SU(2)_{L}\times SU(2)_{R}$ symmetry is realized as follows.
Define the vectors  
$u_1=(a^{(1)},b^{(1)})$, $u_2=(b^{(2)},a^{(2)})$, 
$v_1=(a^{(1)},b^{(2)})$ and  
$v_2=(b^{(1)},a^{(2)})$.
The elements of $SU(2)_{L}$ 
($SU(2)_{R}$) 
act on $u_{1}$ and $u_{2}$ ($v_{1}$ and $v_{2}$) 
in the fundamental representation.
In particular their U(1) subgroups shift the phases  
\eqn\sul{U(1)_{L}:\ a^{(1)\dagger}\to e^{i\theta}a^{(1)\dagger},\ \ \ 
 b^{(1)\dagger}\to e^{-i\theta}b^{(1)\dagger},\ \ \ 
a^{(2)\dagger}\to e^{-i\theta}a^{(2)\dagger},\ \ \ 
 b^{(2)\dagger}\to e^{i\theta}b^{(2)\dagger},} 
\eqn\sul{U(1)_{R}:\ a^{(1)\dagger}\to e^{i\theta}a^{(1)\dagger},\ \ \ 
 b^{(1)\dagger}\to e^{i\theta}b^{(1)\dagger},\ \ \ 
a^{(2)\dagger}\to e^{-i\theta}a^{(2)\dagger},\ \ \ 
 b^{(2)\dagger}\to e^{-i\theta}b^{(2)\dagger}.} 
These U(1) subgroups correspond to 
the shift of axion $C_{L,R}$ in left and
right-moving sectors, respectively. 
They shift the relative phase of particles 
on each side of the Fermi sea.  

The spacetime T-dualities are defined by $g_L:(C_{L},C_{R})
\to (-C_{L},C_{R})$ and $g_R:(C_{L},C_{R})\to (C_{L},-C_{R})$.
The action of $g_L$ ($g_R$), which is included in 
the original $U(4)$ symmetry, is given by
the exchange of $a^{(1,2)\dagger}$ ($a^{(1,2)\dagger}$) 
with $b^{(1,2)\dagger}$ ($b^{(2,1)\dagger}$). 
Note that $g_R$ is exactly the 
same as $(-1)^{F_L}$ defined by \ztwofb.

The full $SU(2)$ generators are given by 
\eqn\sung{J^{a}_L=u_1^\dagger T^a u_1+u_2^\dagger T^a u_2,
\ \ \ \ J^{a}_R=v_1^\dagger T^a v_1+v_2^\dagger T^a v_2,}
where $T^a$ are the Pauli matrices. We choose $a$ such that $a=3$ 
corresponds to the $U(1)$ subgroup realizing the shift symmetry of the axion 
field.
For example,
$J^{3}_{L,R}$ is given by\foot{
We can also define the current which corresponds to the tachyon field 
shift as $J^{0}=a^{(1)\dagger}a^{(1)}+a^{(2)\dagger}a^{(2)}
-b^{(1)\dagger}b^{(1)}-b^{(2)\dagger}b^{(2)}$. 
This is proportional to the fermion number operator. 
The expectation value of 
$\la J^{0} \lb$ is proportional to $\mu$. This generator changes its
sign by the T-duality action $g_{L,R}$.
The Hamiltonian is given by
$H=a^{(1)\dagger}a^{(1)}+a^{(2)\dagger}a^{(2)}
+b^{(1)\dagger}b^{(1)}+b^{(2)\dagger}b^{(2)}$.}
\eqn\jtn{\eqalign{J^{3}_L&=a^{(1)\dagger}a^{(1)}-a^{(2)\dagger}a^{(2)}
-b^{(1)\dagger}b^{(1)}+b^{(2)\dagger}b^{(2)}, \cr  
J^{3}_R&=a^{(1)\dagger}a^{(1)}-a^{(2)\dagger}a^{(2)}
+b^{(1)\dagger}b^{(1)}-b^{(2)\dagger}b^{(2)}.}}
It is easy to see that if we want to have non-zero expectation values for 
these, we have to deform the Fermi sea on the 
left and right in an asymmetric 
way.
These currents, attaining non zero expectation values, 
describe the RR flux 
backgrounds $F_{\pm}\neq 0$. This is consistent with the identification
\fieldt.   

Now we are interested in the other generators $J^{1,2}$, which 
correspond to the `exotic RR fields' $\ti{C}$ and $\ti{\ti{C}}$.
They are explicitly given by
\eqn\defnij{\eqalign{J^{1}_L &=a^{(1)\dagger}b^{(1)}
+b^{(1)\dagger}a^{(1)}+b^{(2)\dagger}a^{(2)}
+a^{(2)\dagger}b^{(2)}, \cr  
J^{2}_L &=-ia^{(1)\dagger}b^{(1)}
+ib^{(1)\dagger}a^{(1)}-ib^{(2)\dagger}a^{(2)}
+ia^{(2)\dagger}b^{(2)},\cr
J^{1}_R &=a^{(1)\dagger}b^{(2)}
+b^{(2)\dagger}a^{(1)}+b^{(1)\dagger}a^{(2)}
+a^{(2)\dagger}b^{(1)}
, \cr  J^{2}_R &=-ia^{(1)\dagger}b^{(2)}
+ib^{(2)\dagger}a^{(1)}-ib^{(1)\dagger}a^{(2)}
+ia^{(2)\dagger}b^{(1)}. \cr}}
We argue that the $SU(2)$ rotated field strengths correspond to 
the expectation values of
 these currents $F_{\pm}=\la J^3_{L,R}\lb$,
$\ti{F}_{\pm}=\la J^1_{L,R}\lb$ and $\ti{\ti{F}}_{\pm}=\la J^2_{L,R}\lb$ 
up to a constant.
Notice that this 
identification is uniquely determined once we fix the interpretation of 
$U(1)$ subgroup (shift of $C$). Under the two T-dualities the currents are 
rotated as follows $J^1 \to J^1$, $J^2 \to -J^2$, and $J^3 \to -J^3$ for 
the left and right sectors respectively.

We can show the $SU(2)_{L}\times SU(2)_{R}$ algebra
\eqn\sut{[ J^a_L, J^b_L ]=2i\ep^{abc} J^{c}_{L},\ \ \
[ J^a_R, J^b_R ]=2i\ep^{abc} J^{c}_{R},\ \ \
[ J^a_L, J^b_R ]=0} 
explicitly.
Let us denote the linear dilaton background $\mu=f=0$ by $|0 \lb$.
Then it is trivial to see that 
\eqn\symsu{J^a_{L,R}|0 \lb=0.} 

Next let us consider the Fermi sea of Fig.3. On a first guess, 
we may conclude that it describes a state such that
only $\la J^3_{L} \lb$ (i.e. $F_{+}$) is
non-zero. Naively this seems to contradict our previous
conjecture that the configuration of Fig.3 can be regarded as a background 
with 
$F_{+}$ and $\ti{F}_{-}$. However, as we will argue below, we can also
consider a mixed state with the right property. 
Note also that since this configuration
is invariant by the action $g_{R}$, the tachyon background field is set to
zero.

Thus it will be very interesting to ask what kind of configuration
of the 
Fermi sea has a non-zero value of the RR flux $\ti{F}_{-}$ 
(or $\la J^1_{R} \lb$).
To find the answer, we define a new basis of creation operators
\eqn\newcr{c_{\pm}^{(1)\dagger}=\f{1}{\s{2}}(a^{(1)\dagger}
\pm b^{(2)\dagger}),\ \ 
c_{\pm}^{(2)\dagger}=\f{1}{\s{2}}(a^{(2)\dagger}\pm b^{(1)\dagger}).}
Then in this basis $J^1_{R}$ is diagonal
\eqn\jod{J^{1}_R=\left(c_{+}^{(1)\dagger}c_{+}^{(1)}+
c_{+}^{(2)\dagger}c_{+}^{(2)}-c_{-}^{(1)\dagger}c_{-}^{(1)}-
c_{-}^{(2)\dagger}c_{-}^{(2)}\right).}

States for which the flux $\ti{F}_{-}$ is non-zero, should be
asymmetric with respect to the fermion number of $c_{+}$ relative to 
that of $c_{-}$. One non-trivial example is the following mixed state
\eqn\mix{|f \lb =\prod_{i=1}^{N_f}c_{+i}^{(1)\dagger}|0\lb
=\prod_{i=1}^{N_f}
\left(\f{a_i^{(1)\dagger}+ b_i^{(2)\dagger}}{\s{2}}\right)|0\lb.}
Such a state has non-zero 
expectation value\foot{The expectation value 
of $J^0$ in this state is zero consistent 
with the fact that the configuration in Fig.3 
describes a state with $\mu=0$.} $\ti{F}_{-}=\la J^1_{R} \lb =f$.
Notice that it is a $g_{R}=(-1)^{F_L}$ invariant mixed state 
of fermions and holes following
the binomial distribution.
If we take the large $f$ limit, this mixed state can be
well approximated by a state with equal fermion
and hole numbers $\f{f}{2}$. In this case 
we obtain the non-zero expectation values $F_{+}=\ti{F}_{-}=f$ with all
other fluxes set to zero as we wanted.

In conclusion we can regard the thin `tail' in the right part
of Fig.3 as the mixed state \mix. Thus this 
Fermi sea represents a background with flux $F_{+}=\ti{F}_{-}=f>0$.
This statement is clear in the large $f$ limit (weak coupling limit), 
where the quantum mixing can be treated classically. This mixing
issue seems to be consistent with the fact that we have D-instanton
corrections near the top of potential.

Finally, we would like to mention that
a similar $SU(2)$ symmetry arises even non-perturbatively.
To see this let us consider the exact creation operator of fermions
for each energy level (classified in terms of even and odd wave functions) 
in the quantum mechanics \foot{Note that the exact potential is
a double well parity invariant potential.}. 
In this formulation we still have the
symmetry mixing fermions and holes as in our previous 
discussion.  We leave the spacetime interpretation of this symmetry as
an interesting future problem to consider.


\newsec{Type 0B Spacetime Gravity}

Let us consider in more details the space-time interpretation
of the 0B string backgrounds discussed in the previous section.
It is instructive to start with the effective action of 0B string
theory\foot{Notice, that the normalization of the R-R field $C$
here differs from the standard normalization in 0B theory:
$C \to C/\sqrt{4\pi}$. This choice of conventions is justified by
connection with two-dimensional gravity models \CGHS\ discussed below.} 
\six \
(written in units $2\kappa^2=1$):
\eqn\obaction{
S_{2d} = \int d^{2} x \sqrt{- g} \Big[
e^{-2 \phi} \big( R + 4 (\nabla \phi)^2 + c
- (\nabla T)^2 + {2 \over \a'} T^2 \big)
- {1 \over 2} e^{-2T} (\nabla C)^2 + \ldots \Big].}
As we discussed in the previous section, when the tachyon field
is set to zero (that is $\mu=0$ in the Fermi sea picture)
the system is characterized by a larger symmetry $SU(2)_L \times SU(2)_R$ asympt
otically,
which acts on the RR fields in 0B theory.
Hence, in this section we focus on backgrounds with $T=0$,
and begin by considering the backgrounds which involve only the usual RR field,
$F_{\pm} = \langle J_{L,R}^3 \rangle$ in the notation of the previous section.
It is easy to verify that the condition $T=0$
is consistent with the tachyon equation of
motion as long as a background has the property
\eqn\czero{ (\nabla C)^2=0. }
When this is the case, we can consistently set $T=0$ and
write the action \obaction\ in the familiar form (CGHS model) \CGHS:
\eqn\cghsaction{
S_{2d} = \int d^{2} x \sqrt{- g} \Big[
e^{-2 \phi} \big( R + 4 (\nabla \phi)^2 + 4 \lambda^2 \big)
- {1 \over 2} (\nabla C)^2 \Big],}
where we introduced $\lambda = {\sqrt{c} \over 2}$, {\it cf.} \CGHS.
In the light-cone variables $x^{\pm} = x^0 \pm x^1$, 
the field strengths are
\eqn\ffeilds{F_+ = \p_+ C \quad , \quad F_- = \p_- C.}

Since for the backgrounds with zero tachyon field, $T=0$,
the effective dynamics of 0B theory reduces to that of the CGHS model,
we can use the results of \CGHS\ to write down the most general
solution consistent with \czero. Specifically, in the conformal gauge
\eqn\cgauge{\eqalign{
& g_{+-} = - {1 \over 2} e^{2\rho}, \cr
& g_{++} = g_{--} =0,
}}
the equations of motion look like:
\eqn\eoms{\eqalign{
& e^{-2\phi} \big( 4\p_+ \rho \p_+ \phi
- 2 \p_+^2 \phi \big) + {1 \over 2} F_+^2 =0, \cr
& e^{-2\phi} \big( 4\p_- \rho \p_- \phi
- 2 \p_-^2 \phi \big) + {1 \over 2} F_-^2 =0, \cr
& e^{-2\phi} \big( 2\p_+ \p_- \phi
-4 \p_+ \phi \p_- \varphi - \lambda^2 e^{2\rho} =0, \cr
& - 4\p_+\p_- \phi + 4\p_+ \phi \p_- \phi
+ 2\p_+\p_-\rho + \lambda^2 e^{2\rho} =0, \cr
& \p_+ \p_- C =0. 
}}
The most general solution to the equations \eoms\ can be written
in terms of two free fields, called $u$ and $w$ in \CGHS.
Furthermore, the general solution to the $C$-field equation
of motion consists of two plane waves: one left-moving and
one right-moving. In order to meet the consistency condition \czero,\
we have to allow only one type of modes. This corresponds
to setting $F_+ =0$ or $F_- =0$. Without loss of generality,
we can assume that $F_- = 0$, {\it i.e.}
\eqn\fleft{ C = C (x^+). }
Then, the most general solution to \eoms\ can be expressed in terms
of a single function, $u=u(x^+)$, which is related to the light-like
matter field $C$ via
\eqn\fviau{
F_+^2 \equiv (\p_+ C)^2 = - 2 u''.
}
The function $u(x^+)$ plays the role of the matter stress-energy tensor:
\eqn\uviaff{
u (x^+) = {M_0 \over \lambda} - \int dx^+ \int dx^+ T_{++} (x^+)
= {M_0 \over \lambda} - {1 \over 2} \int dx^+ \int dx^+ F_+^2.
}
The function $u(x^+)$ also determines the metric \cgauge\
and the dilaton
\eqn\fpwave{
e^{2 \phi} = e^{2 \rho} = {1 \over u (x^+) - \lambda^2 x^+ x^- }.
}
This is the most general solution (written in the $w=0$ gauge) that
depends only on $x^+$. Since such solutions satisfy the condition \czero, \
they all represent consistent backgrounds in 0B string theory.

Sometimes, it is convenient to write the general solution \fpwave\
in a different set of coordinates, so that one of the coordinates
is precisely the dilaton, $\phi$.
Since both $\phi$ and $\rho$ in \fpwave\ have simple and universal
dependence on $x^-$, we replace $x^-$ by $\phi$.
(Since the solution has interesting dependence on the second light-cone
variable, $x^+$, it makes sense to leave this variable intact).
Hence, we need to rewrite \fpwave\ in terms of the variables
$(\phi, x^+)$. A straightforward calculation gives
the metric in the linear dilaton variables:
\eqn\newds{ds^2 = {2 \over \lambda^2 x^+} d\phi dx^+
- {1 + (x^+ u' - u) e^{2\phi} \over \lambda^2 x^+} (dx^+)^2.
}

\fig{The time-dependent Fermi sea 
for the 0B flux background with flux $F_{y^+}=fe^{-y^+/2}$.}
{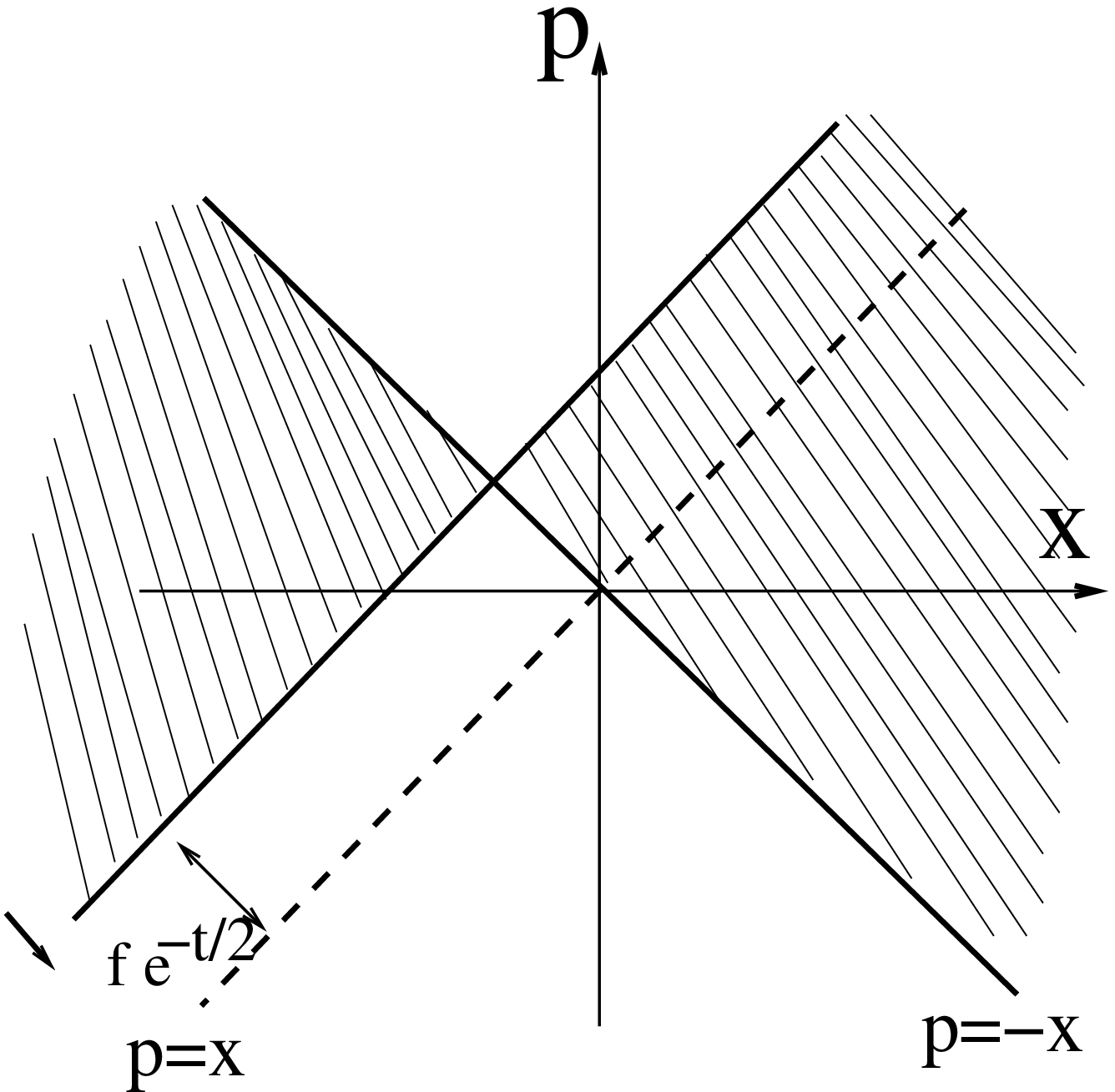}{3.0truein}

Now, let us consider a simple solution
that corresponds to a background with
the light-like R-R flux given by
$F_+ = -\f{f}{\s{x^+}}$. Substituting this flux value into \uviaff, we find
\eqn\solfp{
e^{2 \phi} = e^{2 \rho} = 
-\f{1}{f^2 x^+\log x^+ +\lambda^2 x^+x^-},
}
where, for simplicity, we set $M_0=0$.
Let us define a new set of coordinates
\eqn\cor{x^+=e^{-y^+},\ \ \ f^2\log x^+ +\lambda^2 x^-=-e^{-y^-},}
where $y^+=\phi+t,\ y^-=\phi-t$. Then the background \solfp\
can be written as 
\eqn\backk{ds^2=\f{f^2}{\lambda^2}e^{y^-}(dy^+)^2+\f{1}{\lambda^2}
dy^+dy^- ,} and the dilaton is given by $\phi$.
In these coordinates, the field strength is time-dependent,
$F=fe^{-y^+/2}dy^+$.

It is easy to see that this background describes a space-time,
where particles can freely propagate into the strongly coupled region.
Using \fieldt, we can interpret it as a time dependent Fermi sea (see
\refs{\AKK, \KaSt} for similar discussions
in the two dimensional bosonic string theory)
(see Fig.5) 
\eqn\fermiseatt{\{(x,p)|(p+x)(p-x-fe^{-t/2})<0\}.}

Another important example of a flux background in 0B gravity
corresponds to the Fermi sea \fermisead.
This background describes a static space-time with
two types of fluxes turned on:
\eqn\ffstrange{ F_{+}=f \quad , \quad \ti{F}_{-}=f. }
Notice, that the tachyon tadpole is zero for this background
since the fields $F$ and $\ti{F}$ couple differently to
the tachyon, so that their contributions cancel.
A simple way to see this is to note that the background
\ffstrange\ is invariant under the $\Z_2$ symmetry \ztwofb.
Hence, one can consistently set $T=0$.
The stress-energy tensor for the fields \ffstrange\ is given by
\eqn\emst{T_{++}=\f{1}{2}F_{+}^2=\f{f^2}{2},\ \ 
\ T_{--}=\f{1}{2}\ti{F}_{-}^2=\f{f^2}{2}.}
Therefore, this system is again formally equivalent
to the CGHS system with non-zero flux $F_{\phi}=f$.
The corresponding solution is given by
\eqn\solfpt{e^{2\phi}=e^{2\rho}=-\f{1}{a-f^2\log|x^+x^-|+x^+x^-}.}
After a coordinate change
\eqn\cor{x^+=-e^{-y^+},\ \ \ x^-=e^{-y^-},}
we obtain the metric
\eqn\backk{ds^2=\f{dy^+dy^-}{1-(a+2f^2\phi)e^{2\phi}},}
and the string coupling
\eqn\solfsts{g_s^2=\f{e^{2\phi}}{1-(a+2f^2\phi) e^{2\phi}}.}
Notice, that if we set $f=iq/\s{8\pi}$ and $a=m$,
we obtain the T-dual \solb\ -- \strc\  of the extremal 0A black hole.

Now let us examine the tachyon equation of motion in this spacetime.
Using a simple field redefinition, it is easy to show
that the low-frequency tachyon modes obey a Shr\"odinger-like
equation \schrod\ with the effective potential, $V_{{\rm eff}}$,
which in the asymptotic region looks like
\eqn\shcb{
V_{{\rm eff}} \simeq -(a+2f^2+2f^2\phi)e^{2\phi}.
}
%
As expected from the matrix model,
tachyons are repelled from the strongly coupled region.
Finally, we point out that the $f$-dependence of the time delay
can be estimated as in 0A case, and leads to the result, $f^{-iP}$,
consistent with the matrix model expectation.


\newsec{Type IIB String in Two Dimensions and the Matrix Model Dual}

As we have seen the Fermi sea of Fig.3 is invariant under the $\Z_2$
transformation \ztwofb, 
which can be regarded as particle-hole duality.
Thus we can orbifold the 0B theory with RR-flux by this 
operation. Since this is equivalent to imposing a GSO projection, 
we expect that 
the projected matrix model describes type IIB theory with RR-flux.   
In the end we find a single Fermi surface as shown in Fig.6.

\fig{Fermi sea for Type II string.}
{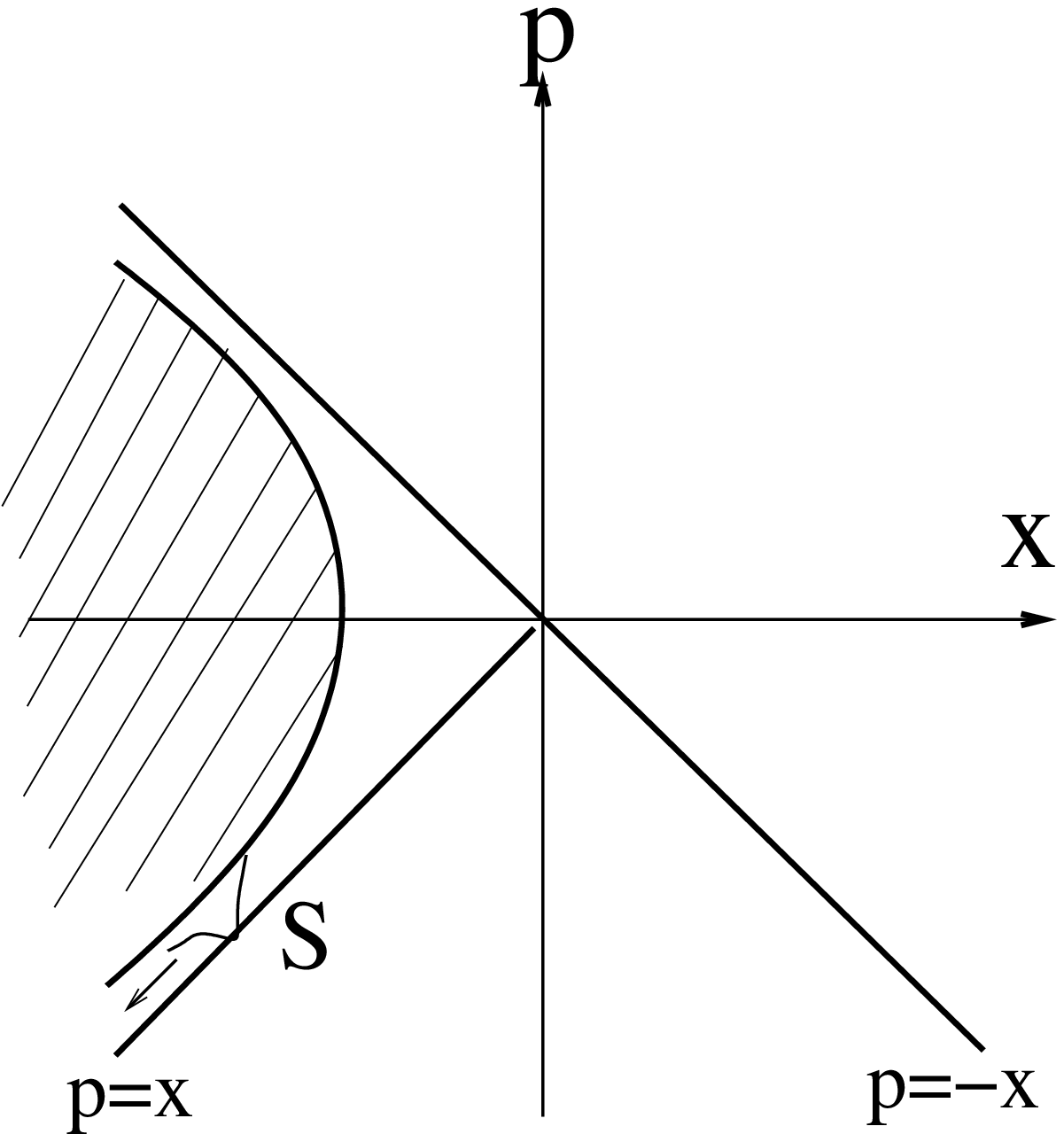}{3.0truein}

Thus we conclude that the projected Fermi sea
describes type II string in two dimensions. 
The field content of this theory\foot{
We can impose the locality condition (GSO projection) as follows 
(see e.g.\TT).
The chiral OPE gives ($\ep$ is the chirality in 2D)
$V_{NS}(z)V_{NS}(0)\sim z^{-1},\ \ 
V_{NS}(z)V_{R(\ep)}(0)\sim z^{-\f12}$ and 
$V_{R(\ep)}(z)V_{R(\ep')}(0)\sim z^{\f{\ep\ep'-1}{4}}$, where we have 
omitted the part which depends on $\phi$ and $X^0$. Thus to maintain the
locality we should choose $\{(R(-),NS),\ (NS,R(-)),\ (R(+),R(+))\}$ in 
type IIB.} is given by 
a left-moving RR scalar field $C(x^+)$ and a right-moving fermion 
$\psi_{RNS}$. The latter can be bosonized\foot{
This implies that the RR scalar (axion) has the compactification radius 
of a free fermion. This is different from that of the axion in type 0B 
string theory (self-radius).
However, this is consistent with the fact that the RR charge of a 
D-brane also differs by factor $\s{2}$.}
\eqn\bosni{\psi_{RNS}(x^-)=e^{i\hat{C}(x^-)/\s{2}}.} 
Note that the way we get $\hat{C}(x^-)$ here is 
quite similar to the one
to obtain $\ti{C}(x^-)$ from $C(x^-)$ in type 0B case (see \selfr).   

Now we get one regular scalar field $C(x^+)+\hat{C}(x^-)$
and this is consistent with the collective field theory
of the previous Fermi sea picture. 
{}From the Fermi sea picture we can see that the 
perturbative scattering amplitudes of $C(x^+)$ and $\hat{C}(x^-)$
are the same as tachyon scattering amplitudes in the 
usual $c=1$ matrix model.
The string coupling is proportional to the inverse of the RR-flux
$F_{+}=\hat{F}_-=f$; thus the flux regulates the 
strongly coupled region\foot{See \MMV\ for discussions on two dimensional 
type IIB string theory in $N=2$ Liouville theory and its matrix model dual.
Obviously their background considered 
is different from ours.}. Notice also that 
in this background the non-perturbative instability still exists as
in 0B case since the $\Z_2$ identification does not eliminate
tunneling effects.
This is consistent with the fact that type IIB still has D-instantons.
This instability is well suppressed when $f$ is large\foot{Even though the 
Fermi sea looks the same as that in bosonic string, there is an important
difference. After we include tunneling effects, the original Fermi sea 
disappears in the bosonic string case, while the Fermi sea will be 
quantum mechanically mixed in type IIB case.}.

\vskip 0.4in

\centerline{\bf Acknowledgments}
We would like to thank V. Balasubramanian,
J. de Boer, R. Dijkgraaf, T. Eguchi, J. Gomis, P.M. Ho,
A. Kapustin, J. Karczmarek, Y. Matsuo, J. Mcgreevy,
S. Murthy, S. Terashima, C. Vafa, H. Verlinde, X. Yi 
and especially
I. Klebanov, J. Maldacena, S. Minwalla, N. Seiberg and A. Strominger
for valuable discussions.
This work was conducted during the period S.G. served as a Clay Mathematics
Institute Long-Term Prize Fellow. S.G. is also supported in part
by RFBR grant 01-02-17488.
T.T. would like to thank Institute for Advanced Study for its hospitality.
The work of T.T. was supported in part by DOE grant DE-FG03-91ER40654.

\listrefs
\end